\documentclass[prf,aps,onecolumn]{revtex4}
\usepackage{graphicx}
\usepackage[usenames]{color}
\usepackage{amsmath,amssymb}
\usepackage{gensymb}
\usepackage{natbib}
\usepackage{longtable}
\usepackage{upgreek}
\usepackage{hyperref}
\usepackage{subfigure}

\newcommand{\bs}{\boldsymbol}

\begin{document}

\title{Transport of flexible fibers in confined micro-channels}
\author{Jean Cappello$^{1,*}$, M. Bechert$^{3,*}$, Camille Duprat$^{2}$, Olivia du Roure$^{1}$,  F. Gallaire$^{3}$ and Anke Lindner$^{1}$}
\affiliation {$^{1}$ Laboratoire PMMH-ESPCI Paris, PSL Research University, 10, rue Vauquelin, Paris, France.\\
$^{2}$ Laboratoire d'Hydrodynamique (LadHyX), \'Ecole polytechnique, Department of Mechanics, Palaiseau, France\\
$^{3}$ Laboratory of Fluid Mechanics and Instabilities, \'Ecole Polytechnique F\'ed\'erale de Lausanne, Lausanne 1015, Switzerland}
\begin{abstract}
When transported in confined geometries rigid fibers show interesting transport dynamics induced by friction with the top and bottom walls.  
Fiber flexibility causes an additional coupling between fiber deformation and transport and is expected to lead to more complex dynamics. 
A first crucial step for their understanding is the characterization of the deformed fiber shape. Here we characterize this shape for a fiber transported in a confined plug flow perpendicular to the flow direction using a combination of microfluidic experiments and numerical simulations. In the experiments, size, initial orientation, and mechanical properties of the fibers are controlled using micro-fabrication techniques and in-situ characterization methods.The numerical simulations use modified Brinkman equations as well as full 3D simulations. We show that the bending of a perpendicular fiber results from the force distribution acting on the elongated object and is proportional to the elasto-viscous number, comparing viscous to elastic forces. We quantitatively characterize the influence of the confinement on the fiber deformation. The precise understanding of the deformation of a flexible fiber in a confined geometry can also be used in future to understand the deformation and transport of more complex deformable particles in confined flows, as for example vesicles or red blood cells. 
\end{abstract}
\date{\today}
\maketitle

\section{Introduction}


The transport dynamics of fibers is important in numerous situations, as the paper-making industry \cite{Stockie1998, Lundell2011}, the clogging of arteries or stents with biofilm streamers \cite{Drescher2013}, the transport of motile micro-organisms \cite{Lauga2009}, filtration devices, or fiber optics used as in situ probes for flows in natural rocks \cite{Dangelo2009}. Transport dynamics of rigid fibers at low Reynolds number has been extensively studied in unbounded media. It is a classical result that rigid fibers drift laterally when sedimenting in a quiescent fluid due to the drag anisotropy of elongated objects in Stokes flow \cite{Batchelor1970}. When transported in shear flows they perform well-known Jeffery orbits \cite{Jeffery1922} while following the stream lines. The presence of confining walls modifies this transport dynamics. It has recently been shown that the transport of rigid fibers in confined geometries shares some of the features observed in sedimentation, notably a lateral drift  \cite{Nagel2018, Berthet2013}, but in the opposite direction compared to sedimentation. Subsequent interaction with lateral walls leads to a rich transport dynamics, including periodic oscillations between the walls \cite{Nagel2018, Uspal2012, Shen2014}. Particles of asymmetric shapes have been shown to reorient with respect to the flow direction \cite{Bet2017, Bet2018} and to reach in specific cases an equilibrium position in the center of the channel \cite{Uspal2012, Shen2014}.\\
On the other hand, the transport dynamics of flexible fibers results from a coupling between deformation and transport and exhibits very rich dynamics. In shear flows this results in a combination of rotation and deformation and fiber buckling can occur \cite{Becker2001, Liu2018}. In more complex flows, deformation induces cross streamline migration \cite{Quennouz2015}. Flexible fibers settling in a quiescent fluid deform, drift and reorient as they sediment until they reach an equilibrium position independent from their initial configuration \cite{Cox1970, Xu1994, LI2013, Marchetti2018}. At equilibrium, the fiber remains horizontal and there is no lateral drift. The transport dynamics of confined flexible fibers, however, is nearly unexplored \cite{DAngelo2013}, but even richer dynamics can be expected as a consequence of the interaction between the fiber, of evolving shape, and the confining walls.\\
\begin{figure}
  \begin{center}
   \includegraphics[width=0.9\columnwidth]{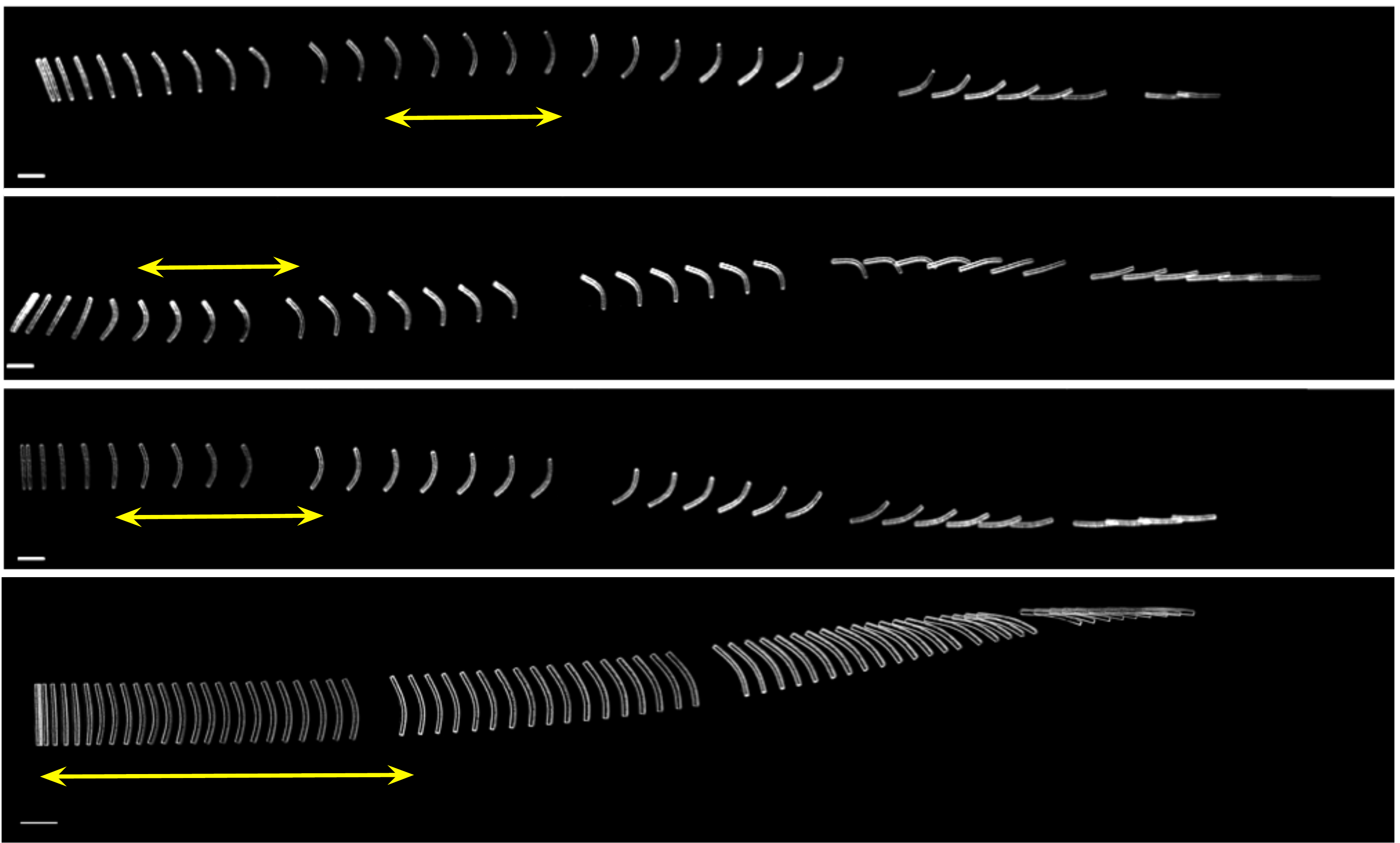}
  \end{center}
  \caption{Superposition of pictures of a transported and deformed flexible fiber for different initial orientations. The height of the channel is 65 $\mu$m and the confinement is around 0.81. The length of the fiber is 800 $\pm$ 10 $\upmu$m, its height is 54 $\pm$ 3 $\upmu$m and its width is 57 $\pm$ 3 $\upmu$m. Scale bar is 500 $\upmu$m long. The mean velocity of the flow is 0.62 mm$\cdot$s$^{-1}$. The yellow arrow indicates the bent state of a deformed fiber perpendicular to the flow. }
\label{fig:reorientation}
\end{figure}
An example of such transport dynamics is shown in Fig. \ref{fig:reorientation} for a microscopic fiber transported in a pressure driven flow in a Hele-Shaw like geometry. The fiber is strongly confined by the top and bottom walls (fiber width and height are comparable to the channel height), whereas the channel width is much larger than the fiber length. The fiber is observed to deform while transported downstream, in particular when oriented perpendicular with respect to the flow direction. This deformation of a freely transported fiber at low Reynolds number in a homogeneous plug flow is at first glance surprising. However, due to friction with the top and bottom walls, the fiber acts as a moving obstacle when pushed down the channel, leading to a strong flow perturbation \cite{Nagel2018}. Despite the fact that the total force acting on the fiber is equal to zero, the deformation indicates that there is a non-homogeneous force distribution along the fiber. During downstream transport there is a coupling between fiber deformation and reorientation. Depending on the initial orientation of the fiber, different dynamics are observed, involving in most cases a deformation to a bent state where the fiber is close to a perpendicular orientation with the flow as highlighted by the yellow arrows in Fig. \ref{fig:reorientation}. This perpendicular orientation is not stable and in all cases the fiber finally aligns with the flow and is advected downstream without deformation or rotation. Reaching the final equilibrium position involves in some cases oscillations of the fiber with respect to the lateral walls. A first step in understanding these complex dynamics is to understand the fiber deformation occurring for a fiber oriented perpendicular to the flow, and in particular to identify the mechanisms leading to a non-homogeneous force distribution responsible for the observed shape.\\
In this paper we investigate this issue in detail with a combination of well-controlled microfluidic experiments and numerical simulations using modified Brinkman equations \cite{Nagel2015, Nagel2018} as well as full 3D simulations. The structure is as follows. We start with an introduction of the terminology and deduce an elasto-viscous number as the primary control parameter in Sec.~\ref{sec:scaling_arg}. The experimental setup and procedure are then described. We control the shape, orientation, and mechanical properties of our particles using micro-fabrication techniques \cite{Berthet2013, Dendukuri2008, Dendukuri2009, Dendukuri2007, Helgeson2011, Lindner2014} and in-situ characterization methods \cite{Wexler2013, Duprat2014} (sec. \ref{sec:expmethods}). In Sec.~\ref{sec:Model}, we present the 3D and 2D model equations employed to calculate the flow around the fiber, the derivation of the fiber deflection, and the numerical methods. The experimental and theoretical results are finally shown and discussed in Sec.~\ref{sec:results}. For the first time, we explore how the local distribution of the viscous force along the fiber determines its deformation. We systematically investigate fiber shape and fiber deflection as a function of fiber geometry and confinement and show that the amplitude of bending is proportional to the elasto-viscous number and increases with the confinement. The paper closes with a brief conclusion and outlook.

\section{Physical mechanisms, scaling arguments and geometrical arrangement}
\label{sec:scaling_arg}

\begin{figure}[t]
	\begin{center}
		\includegraphics[width=.9\columnwidth]{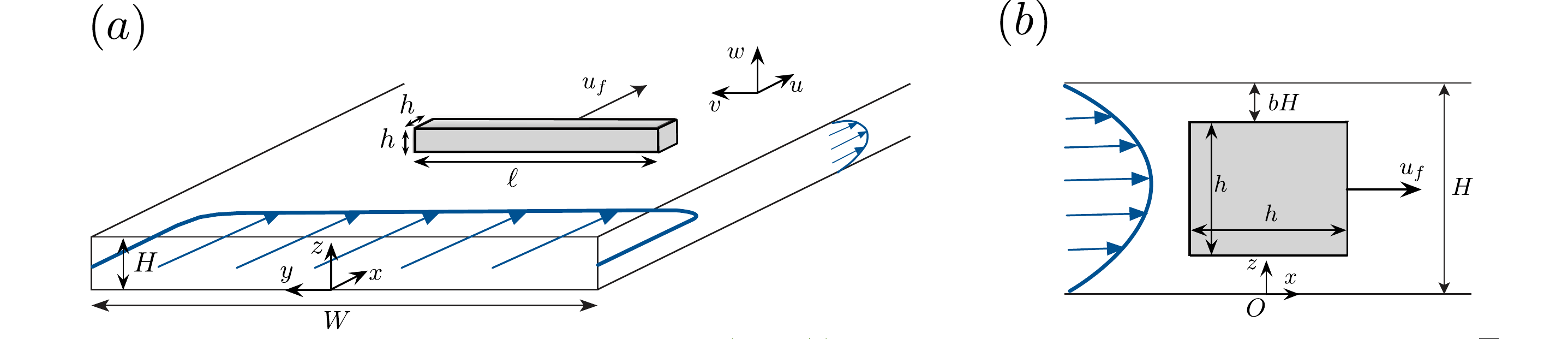}
	\end{center}
	\caption{(a) Geometry of the perpendicular fiber and of the channel. The pressure-driven flow is sketched in blue and defined by a mean velocity $u_0$, while the fiber is moving with a velocity $u_f$ so that the total force on the fiber surface becomes zero.\\(b) Cross section view of the channel and of a perpendicular fiber. The height of the gap $bH$ is determined by the confinement $\beta=1-2b$.}
	\label{fig:normalfor_plus_f_lat}
\end{figure}

The shape of an elastic fiber interacting with a viscous flow is given by a balance between drag, i.e. pressure and viscous forces, and elastic restoring forces \cite{DuRoure2019}. Their balance is expected to determine the fiber shape. In the specific situation of a freely transported fiber in a viscous flow the total force acting on the fiber is zero and fiber deformation can only occur due to a non-homogeneous force distribution acting on the fiber. This force distribution can for example result from the straining part of a shear flow. Here, we study the deformation of a single flexible fiber transported in a confined channel (see Fig. \ref{fig:normalfor_plus_f_lat} (a) and (b)) where the fiber is transported in a plug flow. In this case the viscous and pressure force distributions result from the disturbance flow occurring around the fiber, which thus need to be determined.

The precise geometry considered is the following. The fiber is an elongated object of square cross-section with width $h$ and length $\ell$ so that the aspect ratio $\rm ar=\ell/\it h$ is large (19 to 25). The channel is rectangular, of constant height $H$, width $W$ and length $L$ with $W,L \gg H$. The fiber is located at the center of the channel. The confinement $\beta=h/H$ quantifies the influence of the top and bottom channel walls on the fiber dynamics and the height of each gap above and below the fiber is given by $bH$, with $2b=1-\beta$. The fiber is transported by an external pressure-driven flow characterized by the mean velocity $u_0$. The velocity field is denoted by $\bs u=(u,v,w)$, with $u$, $v$ and $w$ being the velocity components in $x$, $y$ and $z$ direction, respectively, and the pressure field is denoted by $p$. The fluid is a Newtonian liquid with constant shear viscosity $\mu$. Due to the Hele-Shaw geometry of the channel, the flow in the $xy$-plane is a plug flow, except close to the side walls and in the vicinity of the fiber. For the flexible fibers treated here, we restrict our analysis to fibers oriented perpendicularly to the flow direction, corresponding to the state of maximum deformation as observed experimentally (Fig.~\ref{fig:reorientation} and Sec.~\ref{sec:exp_fib_transp}).\\
The fiber shape is determined by first evaluating the flow around a rigid fiber. The transport velocity of a rigid fiber is given by the balance between the pressure and viscous forces pushing the fiber and the viscous friction occurring in the small gaps between the moving fiber and the fixed top and bottom walls. The equilibrium velocity $u_f$ is then obtained by imposing a force-free condition on the fiber surface. Note that all previous analyses considering rigid fibers in confined geometries have only investigated total force and torque balances \cite{Champmartin2010, DAngelo2013,Berthet2013, Nagel2018}. However, as mentioned above, for a flexible, perpendicular fiber the inhomogeneity of the force along the fiber is crucial for the description of the deformation. We thus need to determine the force distribution per unit length along the fiber $f(y)$ resulting from pressure and viscous friction. Finally, the fiber deflection $\delta$ caused by this force distribution is calculated employing the linear Euler-Bernoulli equation, given by
\begin{align}
EI\,\frac{\partial^4 \delta}{\partial y^4} = f(y),
\label{eq:Euler_Bernoulli}
\end{align}
with flexural rigidity $E I$, $E$ being the material Youngs modulus and $I=h^4/12$ being the areal moment of inertia. For an homogeneous slender rod, the flexural rigidity is independent of $y$. Resolving this equation requires the use of appropriate boundary conditions, which will be discussed in Sec.~\ref{sec:bc}. Note that we assume the effect of fiber bending on the flow to be negligible throughout this paper, as the observed deflections are small.
\\
We scale $x$, $y$, $z$ and $\delta$ by the fiber length $\ell$, the velocity components by $u_0$, the pressure and stress tensor components by $\mu\,u_0/\ell$ and the forces per unit length $f$ consequently by $\mu\,u_0$. All dimensionless variables are denoted by a tilde. This scaling leads to the dimensionless geometry
\begin{align}
\label{eq:scaled_lengths}
\tilde\ell = 1,\ \ \ \tilde h = \frac{1}{\rm ar} \ \ \ \text{and}\ \ \ \tilde H=\frac{1}{\rm ar\,\beta}.
\end{align}
Equation~(\ref{eq:Euler_Bernoulli}) transforms to
\begin{align}
\label{eq:Euler_Bernoulli_scal}
\frac{\partial^4 \tilde\delta}{\partial \tilde y^4} = \tilde\mu\,\tilde f(\tilde y),
\end{align}
with
\begin{align}
\label{eq:mutilde}
\tilde\mu = \frac{\mu\,u_0\,\ell^3}{EI},
\end{align}
referred to as the elasto-viscous number \cite{DuRoure2019, Wandersman2010, Quennouz2014}. With the drag force on the fiber being proportional to $u_0$, $\tilde\mu$ compares viscous drag to elasticity, which are the main forces controlling the dynamics provided that the effect of inertia and Brownian motion is negligible. According to Eq.~(\ref{eq:Euler_Bernoulli_scal}), the scaled fiber deflection is expected to depend linearly on the elasto-viscous number. While the drag force is proportional to the mean velocity of the surrounding fluid, its exact form and amplitude $f(y)$ also depend on the confinement $\beta$. The dependence of the fiber deformation on~$\tilde\mu$ and $\beta$ is one of the main questions investigated in the rest of this paper.
%
%
%
%
\section{Experimental methods}
\label{sec:expmethods}
\subsection{Channel and fiber fabrication}
\label{subsec:fabrication}
The microchannels are polydimethylsiloxane (PDMS) channels produced from molds fabricated using a micro-milling machine (Minitech Machinery). The channels are bonded to a cover slide spin coated with a thin layer of PDMS in order to ensure identical boundary conditions on the four walls. We use straight channels of width $W$=3500 $\upmu$m with varying heights $H$ from 30 to 85 $\upmu$m and a length of a few centimeters. We have verified that the channel height does not vary more than 3 $\upmu$m along the channel width or length, even if the aspect ratio $W/H$ is very large .

\begin{figure}
	\begin{center}
		\includegraphics[width=1\columnwidth]{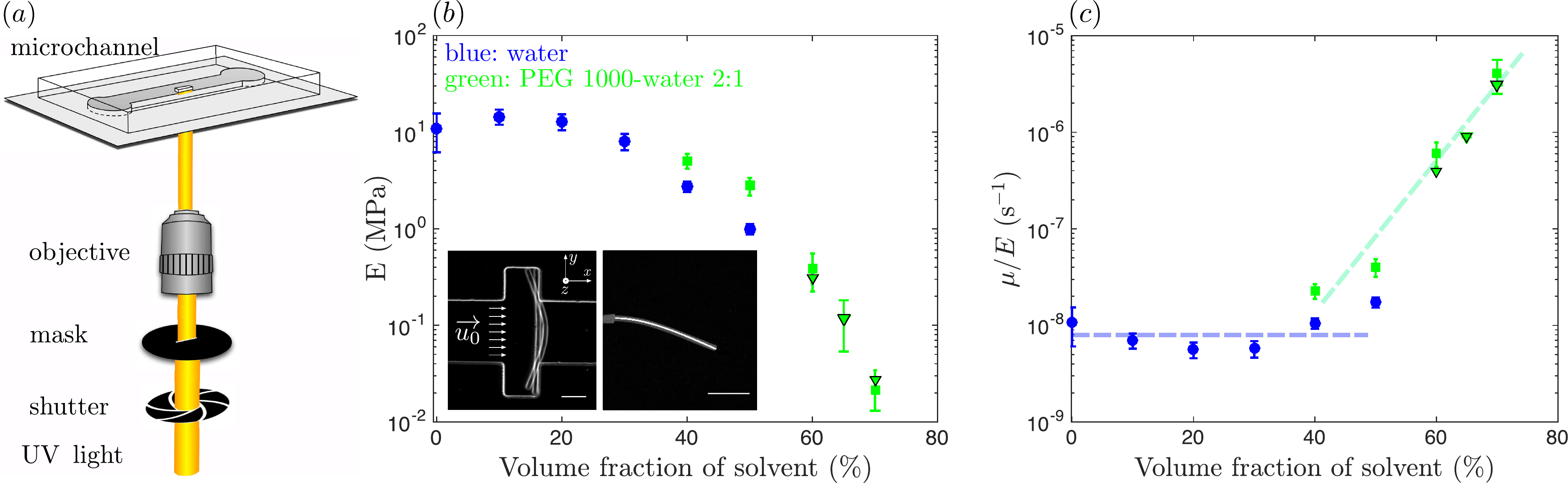}
	\end{center}
	\caption{(a) In situ fabrication of the fiber using a microscope-based projection lithography technique. UV light is projected into the channel through a fiber-shaped mask, polymerizing the photosensitive solution inside the channel.  \\(b) and (c): Characterization of the Youngs modulus of the fibers. Blue and green symbols correspond respectively to dilution with water and with a solution of PEG 1000 - water (2/3-1/3 in volume). Measurements were obtained using  the in-situ method \cite{Duprat2014}, and completed with macroscopic measurement (triangles).\\
(b) Evolution of the Youngs modulus as a function of the percentage of dilution.  Bottom left inset shows the in situ mechanical characterization of the Youngs modulus of the fiber using the technic developed by Duprat \textit{et al.} \cite{Duprat2014}. Scale bar is 100 $\upmu$m. Bottom right inset corresponds to a macroscopic measurement of the Youngs modulus using the shape of a fiber deformed by its own weight. The white line is a fit with the theoretical shape. Scale bar is 1 mm.\\(c) Evolution of the ratio of viscosity to Youngs modulus as a function of dilution. While the change in viscosity cancels the influence of solvent fraction on $\mu/E$ for a dilution with water, the quantity can be tuned within several orders of magnitude for a dilution with PEG 1000-water.}
	\label{fig:E_pcofwater_orPEG1000+water}
\end{figure}

We fabricate fibers of controlled geometry using the stop-flow microscope-based projection photo-lithography method developed by Dendukuri \textit{et al.} \cite{Dendukuri2008} presented in Fig.\ref{fig:E_pcofwater_orPEG1000+water} (a) and used previously for the investigation of rigid fibers by Berthet \textit{et al.} \cite{Berthet2013}. The microfluidic channel is filled with a solution of oligomer and photo-initiator and exposed to a pulse of UV light through a lithography mask placed in the field-stop position of the microscope (see Fig. \ref{fig:E_pcofwater_orPEG1000+water} (a)). We work with a Zeiss Axio Observer equipped with a UV light source (Lamp HBO 130W) and a X5 Fluar objective. The exposure time is precisely controlled using an electronic shutter (V25, Uniblitz) coupled to an external generator (Agilent 33220A). Due to the permeability of PDMS to oxygen inhibiting the polymerization, a non-polymerized lubrication layer of constant thickness is left along the top and bottom walls of the channel \cite{Dendukuri2008}. Due to these lubrication layers the fabricated fibers can be transported in the channel.

The polymeric fibers are fabricated at zero flow rate with a given orientation at the center of the channel to avoid interactions with the lateral walls. The fiber length $\ell$ and width $h$ are determined by the mask shape and the specific properties of the microscope and objective used \cite{Berthet2016}. The maximum fiber length that can be obtained with this set-up is 1500~$\upmu$m. The height of the inhibition layer is found to be of constant thickness $(H -h)/2 = 5.5 \pm$1.6 $\upmu$m in our setup  \cite{Berthet2013, Wexler2013, Duprat2014, Nagel2018}. The fiber height $h$ as well as the confinement $\beta=h/H$ are thus both solely determined by the height $H$ of the channel. The channel and the fiber height are measured with an error of $3~\upmu$m leading, for confinements varying from 0.86 to 0.6, to uncertainties varying from 0.04 to 0.1. We adjust the mask shapes as a function of the channel height to ensure a square cross-section and an aspect ratio of $\ell/h = 21$ or 25. Typical fiber dimensions vary from $19~\upmu$m to $74~\upmu$m for width and height and from 0.5 mm to 1.5 mm for the length. The fiber length $\ell$ is always smaller than $0.43\, W$ in order to neglect the influence of the side walls.  We verified experimentally that indeed up to $\ell/W \gtrsim 0.6$ the fiber deflection is independent of the lateral confinement. Only above this threshold the influence of the lateral walls becomes noticeable and the deflection is observed to decrease with increasing confinement (data not shown). The photosensitive solution is composed of polyethyleneglycol-diacrylate (PEGDA, Mw = 700, Sigma) and a solvent at varying proportions. The solvent is either water or a mixture of water and polyethyleneglycol (PEG, Mw = 1000, Sigma) at a ratio of 1:2 in volume. The photosensitive solution always contains 10$\%$ of Darocur 1173 photo-initiator (PI, 2-hydroxy-2-methylpropiophenone, Sigma).
\subsection{Mechanical characterization}
Compared to other situations studied previously (\cite{Duprat2014, Wexler2013, Berthet2016}), where the flexible fiber is held fixed in the channel, the fiber studied here is free and the deformations induced by the fluid are rather small. In order to obtain significant deformation we work with highly flexible fibers, i.e. fibers with large aspect ratio $\rm ar$ or small elastic moduli, thus small flexural rigidity. The aspect ratio is limited by the maximum fiber length possible in our set-up (see sec. \ref{subsec:fabrication}) and we thus attempt to fabricate fibers of low modulus with good accuracy and reproducibility.\\
The Young's modulus can be tuned either by varying the composition of the photosensitive mixture or by adjusting the UV exposure time  \cite{Duprat2014, Chen2017}, the latter influencing the Youngs modulus exponentially until a plateau is reached for large times \cite{Duprat2014}. As the control over the exposure time is not accurate enough even with the high-precision shutter used (V25, Uniblitz) we perform all experiments in the plateau regime using a constant exposure time of 600 ms and we tune the Young's modulus by diluting the PEGDA with water or with a mixture of water and PEG1000.\\
The Youngs modulus is measured using the \textit{in situ} technique developed by Duprat \textit{et al.} \cite{Duprat2014}. Fibers held fixed inside lateral notches are deformed under the action of a viscous flow. The Youngs modulus is then obtained from the shape of the fiber. Details of the method can be found in Duprat \textit{et al.} \cite{Duprat2014}. Here, straight microchannels (cross-section is 200 $\upmu$mx40-70 $\upmu$m, and length is a few centimeters) are designed with regularly spaced slots, (120 $\upmu$m-long). The fibers are fabricated into the notches without flow as illustrated in the bottom right inset of figure \ref{fig:E_pcofwater_orPEG1000+water} (b) and then submitted to an external flow. Fiber dimensions are similar for this {\it in-situ} measurement of the Young's modulus and the experiments on freely transported fibers ($h$=30-40$\upmu$m,  $w$=20$\upmu$m, $\ell$=500 $\upmu$m). However, for highly flexible fibers the deformations are too important and the fiber escapes from the notches. Therefore, we complement our in-situ measurements with macroscopic measurements of the  Young's modulus of a larger fiber deformed by its own weight. For this experiment the photosenitive solution is illuminated with UV light in a tubing of inner diameter 0.64 mm. The crosslinked fiber is then extracted from the tubing by pushing with a syringe. The typical lengths studied span a range from a few millimeters to a few centimeters. The Young's modulus is again obtained from the shape of the deformed fiber \cite{Audoly2000, Quennouz2013}; the bottom left inset of figure \ref{fig:E_pcofwater_orPEG1000+water} (b) shows a deformed fiber and the theoretical adjustment used to obtain the Young's modulus (white line). Figure  \ref{fig:E_pcofwater_orPEG1000+water} (b) shows the Young's modulus of dilutions of PEGDA with increasing percentages of water or of the water/PEG 1000  mixture. Macroscopic and microscopic results are in good agreement. The Young's modulus of the gel decreases as the volume fraction of solvent increases and moduli as low as 10 kPa can be obtained.
In our experiments, the fibers are transported in the uncrosslinked photosensitive mixture. Diluting the photosensitive mixture affects not only the Young modulus of the fiber but also the viscosity of the surrounding fluid. The value of interest is actually the ratio $\mu/E$ as the elasto-viscous number $\tilde{\mu}$ is proportional to $\mu/E$ according to Eq.~(\ref{eq:Euler_Bernoulli_scal}). We plot this ratio as a function of the dilution fraction in Fig. \ref{fig:E_pcofwater_orPEG1000+water}~(c). Diluting PEGDA with water the elastic modulus $E$ of the crosslinked fiber decreases, but the ratio $\mu/E$ remains constant: the gain in flexibility is cancelled out by the loss of viscosity. Diluting PEGDA with the more viscous PEG 1000/water mixture on the contrary leads to a variation of $\mu/E$ of three orders of magnitude allowing for larger deformations. 
\subsection{Fiber transport and image treatment}
\label{sec:exp_fib_transp}
\begin{figure}
	\begin{center}
		\includegraphics[width=.7\columnwidth]{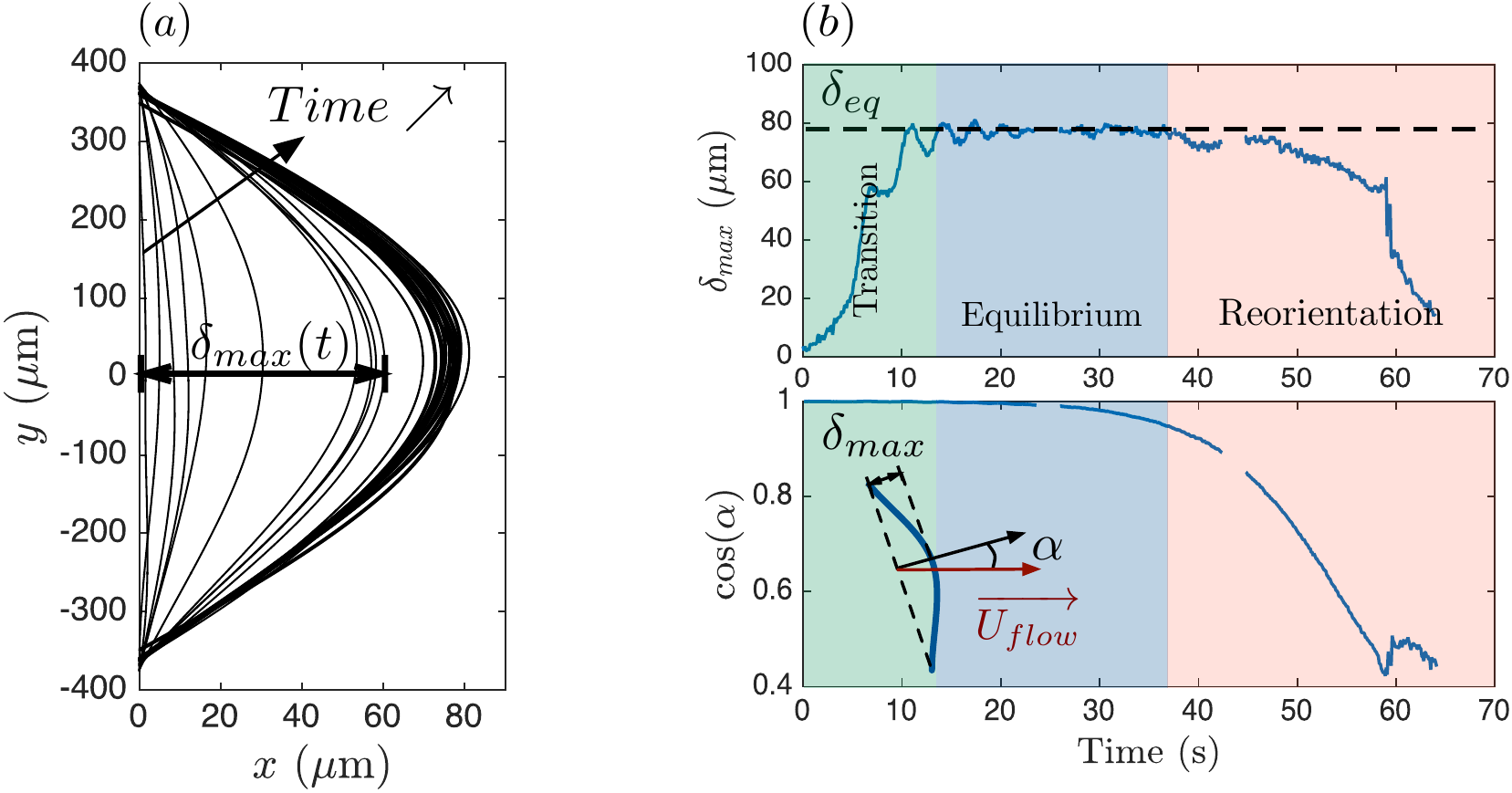}
	\end{center}
	\caption{ Evolution of fiber deformation with time. The data corresponds to the experiment shown on the bottom row of figure \ref{fig:reorientation}. (a) Fiber shape evolution. The time step between two line is $1$s. (b) Top: Evolution of the maximum deflection of the fiber $\delta_{\rm max}$ as a function of time. The equilibrium deflection $\delta_{\rm eq}$ corresponds to the value of the plateau. Bottom: evolution of the angle~$\alpha$ between the fiber and the flow as a function of time. $\cos(\alpha)=1$ corresponds to a perpendicular orientation of the fiber.}
	\label{fig:exp_obs_delta_eq}
\end{figure}
Once the fibers are fabricated inside the channel, flow is turned on and fiber transport is recorded using a X2.5 EC Epiplan-Neofluar objective and a Hamamatsu Orca-flash 4.0 camera at a frame-rate of 10 images per second. A precision pump (Nemesys, Cetoni) drives the flow in the channel. The microscope stage is displaced by hand to keep the fiber in the field of view of the camera. The fiber shape is obtained using standard image treatment procedures (with ImageJ \cite{Schneider2012} and Matlab) and is recorded as a function of time together with the maximum deflection $\delta_{\rm max}$ (Fig.~\ref{fig:exp_obs_delta_eq}~(a)). To measure the fiber deflection of a perpendicular fiber at equilibrium $\delta_{\rm eq}$ we monitor the maximum fiber deflection as well as the fiber orientation as a function of time. Fig.~\ref{fig:exp_obs_delta_eq}~(b) shows that after a transition period fiber deflection and orientation are stable over a given time before the fiber changes orientation and fiber deflection decreases.  At equilibrium, the fiber remains perpendicular to the flow and translates along the x-axis only. Our measurements are always performed in the equilibrium regime.\\
Velocity and pressure distributions can be obtained from streakline pictures (as can bee seen in Fig.~\ref{fig:streaklines}~(a)). They are reconstructed images built from processing several successive image captures of the flow with suspended latex particles of 1 $\upmu$m diameter, all superimposed while keeping the fiber in the center of the image.  The velocity field around the fiber is obtained using particle tracking techniques in the frame of reference of the fiber. Velocity is calculated by averaging the particle velocity in 64$\times$64 pixels windows, and once the system has reached its stationary regime, by averaging over time. The X10 ultrafluar objective used allows us to image the full channel depth, the flow field obtained thus corresponds to the depth-averaged velocity of the fluid.  The pressure field can be obtained integrating the velocity field using the Brinkman equation (\ref{eq:Brikman_eq}) introduced below. The Brinkman equation involves a second derivative of the velocity, so to reduce noise in the pressure field a median filter is used on the velocity signal and a central difference method of second order accuracy (including four points) is applied.\\
%
%
%
\section{Modeling}
\label{sec:Model}
\subsection{3D model}
We start with introducing a 3D~model to calculate the flow around a rigid, perpendicularly oriented fiber and the resulting forces on the fiber surfaces. This model will mainly serve as a reference for the reduced 2D model introduced in Sec.~\ref{sec:2D_model_eqs}. The scaled velocity and pressure fields are determined by the well-known equations for Stokes flow,
\begin{subequations}
\label{eq:stokes}
\begin{align}
\bs\nabla\cdot\tilde{\bs{u}} &= 0,\\
-\bs\nabla\tilde p + \Delta\tilde{\bs u} &= 0.
\end{align}
\end{subequations}
We impose slip conditions on the lateral walls of the channel to simulate an infinite channel width and additionally no-slip conditions on the top and bottom walls. The fiber velocity $\tilde u_f\,\bs{e_x}$ is prescribed on the surface of the fiber. The inlet velocity is defined to have a Poiseuille profile in $z$ with unit mean velocity and a normal flow and constant pressure are specified at the outlet. The forces imposed by the flow on the fiber surfaces are determined by integrating along the $x$ and $z$ axes, leading to forces per unit length in $x$-direction on the fiber front and tail, denoted by $\tilde f_{\rm front}$ and $\tilde f_{\rm tail}$, respectively, as well as to the cumulated force on top and bottom of the fiber $\tilde f_{\rm gap}$ and the total force on the edge surfaces $\tilde F_{\rm e}$. As we are restricting the analysis to fibers with perpendicular orientation, only the force components in $x$-direction are relevant, which are given as
\begin{subequations}
\label{eq:3D_forces}
\begin{align}
\tilde f_{\rm front}(\tilde y) &= -\int_{b\tilde H}^{(1-b)\tilde H}d\tilde z\,\tilde \sigma_{xx}|_{\tilde x=-\tilde h/2},\\
\tilde f_{\rm tail}(\tilde y) &= \int_{b\tilde H}^{(1-b)\tilde H}d\tilde z\,\tilde \sigma_{xx}|_{\tilde x=\tilde h/2},\\
\tilde f_{\rm gap}(\tilde y) &= \int_{-\tilde h/2}^{\tilde h/2}d\tilde x\,\left(\tilde\sigma_{xz}|_{\tilde z=(1-b)\tilde H}-\tilde \sigma_{xz}|_{\tilde z=b\tilde H}\right),\\
\tilde F_{\rm e} &= \int_{b\tilde H}^{(1-b)\tilde H}d\tilde z\int_{-h/2}^{h/2}d \tilde x\,\left(\tilde \sigma_{xy}|_{\tilde y=0.5}-\tilde\sigma_{xy}|_{\tilde y=-0.5}\right),
\end{align}
\end{subequations}
with the stress tensor components
\begin{align}
\label{eq:Newton_const}
\tilde\sigma_{ij} &= -\tilde p\,\delta_{ij}+(\partial_i \tilde u_j+\partial_j\tilde u_i).
\end{align}
The signs in Eqs.~(\ref{eq:3D_forces}) are determined by the orientation of the surface normal vectors. The fiber will move exclusively in $x$-direction with constant speed $\tilde u_f$, which is determined by the condition of zero total force $\tilde F_{\rm tot}$, i.e., 
\begin{align}
\label{eq:zero_force_3D}
\tilde F_{\rm tot}=\int_{-1/2}^{1/2}d\tilde y\,\left(\tilde f_{\rm front}(\tilde y)+\tilde f_{\rm tail}(\tilde y)+\tilde f_{\rm gap}(\tilde y)\right)+\tilde F_{\rm e} = 0.
\end{align}
With the Stokes equations~(\ref{eq:stokes}) being linear, it is possible to determine $\tilde{u_f}$ by superposition of two independent solutions \cite{Nagel2018}. In particular, the flow and the total force are calculated for two configurations, a fiber fixed in a flow $\left\{\tilde u_0^{(1)}=1,\tilde u_f^{(1)}=0\right\}$ and a fiber moving in a quiescent fluid $\left\{\tilde u_0^{(2)}=0,\tilde u_f^{(2)}=1\right\}$. The equilibrium velocity fulfilling the force-free condition~(\ref{eq:zero_force_3D}) can then be calculated to
\begin{align}
\tilde u_f = -\frac{\tilde F_{\rm tot}^{(1)}}{\tilde F_{\rm tot}^{(2)}},
\end{align}
where the superscripts indicate the configuration. 
\subsection{2D depth-averaged model}
\label{sec:2D_model_eqs}
As the calculations for the 3D model require a lot of computational power, we also introduce a refined version of the depth-averaged 2D model proposed by \citet{Nagel2018}. This model is based on the Brinkman equations,
\begin{subequations}
\label{eq:Brikman_eq}
\begin{align}
\bs\nabla\cdot{\bs{\bar u}} &= 0,\\
\label{eq:Brinkman_mom}
\left(\nabla^2{\bs{\bar u}}-\frac{12}{\tilde{H}^2}{\bs{\bar u}}\right)-\bs\nabla \bar p &= 0.
\end{align}
\end{subequations}
We use a bar above the variables to indicate averaging across the height of the channel, i.e.,
\begin{align}
\bs{\bar u} = \frac{1}{\tilde H}\int_0^{\tilde H} dz\,\bs{\bar u}\,\,\,\, \text{and}\,\,\,\,
\bar p = \frac{1}{\tilde H}\int_0^{\tilde H} dz\,\bar p,
\end{align}
and the differential operators correspond consequently to a two-dimensional space. The 2D flow is then calculated by defining a composite particle containing the fiber and the fluid in the gap between the fiber and the top and bottom channel walls. A constant flow velocity in $x$-direction $\bar{u_p}$ is imposed on the surface of the composite particle and slip conditions are set on the lateral walls to simulate an infinite channel width. Moreover, a plug flow profile with unit velocity at the inlet and a normal flow with constant pressure at the outlet are prescribed.\\
In a second step, the forces imposed by the flow on the fiber surfaces are determined. In contrast to the 3D model, only the height-integrated forces per unit length in $x$-direction on the fiber front and tail, $\tilde f_{\rm front}$ and $\tilde f_{\rm tail}$, respectively, and the total force on the edge surfaces $\tilde F_{\rm e}$ are directly given by
\begin{subequations}
\label{eq:2D_forces}
\begin{align}
\tilde f_{\rm front}(y) &= -\tilde h\,\bar\sigma_{xx}|_{\tilde x=-\tilde h/2} \text{\ \ \ for\ \ } -1/2\leq \tilde y \leq 1/2,\\
\tilde f_{\rm tail}(y) &= \tilde h\,\bar\sigma_{xx}|_{\tilde x=\tilde h/2} \ \ \ \ \text{\ \ \ for\ \ } -1/2\leq \tilde y \leq 1/2,\\
\tilde F_{\rm e} &= \tilde h\int_{-\tilde h/2}^{\tilde h/2}d\tilde x\,\left(\bar\sigma_{xy}|_{\tilde y=1/2}-\bar\sigma_{xy}|_{\tilde y=-1/2}\right).
\end{align}
\end{subequations}
In order to evaluate the forces on top and bottom of the fiber, an additional model for the flow in the gap has to be proposed. As we are considering fibers perpendicular to the flow, the composite particle and thus the fiber are moving only in the $x$-direction, and it is possible to extend the gap flow profile as introduced by \citet{Nagel2018} to account for variations along the fiber length $\tilde{y}$. This leads to a total velocity profile of the composite particle $\bs{\bar u_p} =\bar u_p\,q(\tilde y,\tilde z)\,\bs e_x$, where the gap flow profile $q(\tilde y,\tilde z)$ has to fulfill the normalization condition
\begin{align}
\label{eq:gap_flow_norm_cond}
\frac{1}{\tilde H}\int_{-1/2}^{1/2}d\tilde y\int_0^{\tilde H} d\tilde z\,q(\tilde y,\tilde z) = 1.
\end{align}
Following \citet{Nagel2018}, $q(\tilde y,\tilde z)$ is assumed to be of Couette--Poiseuille type in $\tilde z$, but now has an additional $\tilde y$-dependence, i.e.,
\begin{align}
q(\tilde y,\tilde z) = \begin{cases}
q_1(\tilde y,\tilde z) = C_1(\tilde y)\left(\frac{\tilde z}{\tilde H}\right)^2+C_2(\tilde y)\frac{\tilde z}{\tilde H}\ \ \ &\text{for}\ \ \ 0\leq \tilde z\leq b\tilde H,\\
q_2(\tilde y,\tilde z) = \frac{\bar u_f}{\bar u_p}\ \ \ &\text{for}\ \ \ b\tilde H<z<(1-b)\tilde H,\\
q_3(\tilde y,\tilde z) = C_1(\tilde y)\left(1-\frac{\tilde z}{\tilde H}\right)^2+C_2(\tilde y)\left(1-\frac{\tilde z}{\tilde H}\right)\ \ \ &\text{for}\ \ \ (1-b)\tilde H\leq \tilde z\leq b\tilde H,
\end{cases}
\end{align}
where the no-slip condition at the channel walls, $q(\tilde y,0)=q(\tilde y,\tilde H)=0$ was already applied. Note that the gap flow profile presented by \citet{Nagel2018} is identical to the reduced profile $\int_{-1/2}^{1/2}d\tilde y\,q(\tilde y,\tilde z)$. $C_1(\tilde y)$ and $C_2(\tilde y)$ are determined by the no-slip condition assumed on the fiber surface, $q(\tilde y,b\tilde H)=q(\tilde y,(1-b)\tilde H)=\bar u_f/\bar u_p$ and by applying the Stokes equation in the gap, which introduces the pressure gradient $\partial_{\tilde x}\bar p$. Unlike the model proposed by \citet{Nagel2018}, here, $C_1(\tilde y)$ and $C_2(\tilde y)$ are no longer constants and take into account the variation of the pressure gradient along the fiber length. Due to symmetry and for the sake of simplicity, we will restrict the rest of the analysis to the bottom gap flow $q_1$, which can finally be written as
\begin{align}
\label{eq:q1_prof}
q_1(\tilde y,\tilde z) = \frac{1}{\tilde u_p}\left[-\frac{{\tilde H}^2}{2}\,\partial_{\tilde x}\tilde p\left(b-\frac{\tilde z}{\tilde H}\right)\frac{\tilde z}{\tilde H}+\bar u_f\frac{\tilde z}{b\tilde H}\right].
\end{align}
As all forces along the fiber have to be symmetric in $\tilde y$, only their mean value contributes to the movement of the rigid composite particle. The correlation between fiber velocity $\bar u_f$ and composite particle velocity $\bar u_p$ remains thus identical to the one obtained from the model of \citet{Nagel2018} and \citet{Berthet2013} and can be written as
\begin{align}
\label{eq:fib_vel_2D}
\bar u_f &= \frac{3\,(1+\beta)}{2\,(1+\beta+\beta^2)}\,\bar u_p.
\end{align}
The average of $\partial_{\tilde x}\bar p$ along the fiber is implicitly given by Eq.~(\ref{eq:gap_flow_norm_cond}) and can be expressed employing Eq.~(\ref{eq:q1_prof}) by
\begin{align}
\label{eq:px_av}
\langle\partial_{\tilde x}\bar p\rangle = \int_{-1/2}^{1/2}d\tilde{y}\partial_{\tilde x}\bar p = -\frac{12\,\bar u_p}{\tilde H^2(1-\beta^3)}.
\end{align}
The equality in Eq.~(\ref{eq:px_av}) was separately verified for various values of $\beta$ as shown in Appendix~\ref{app:consistency_2D_model} and ensures that the presented model is a consistent extension of the one introduced by \citet{Nagel2018}, where only averaged values of the forces are used. Approximating the pressure gradient along the fiber by
\begin{align}
\label{eq:px_approx}
\partial_{\tilde x} \bar p \approx \frac{\bar p|_{\tilde x=\tilde h/2}-\bar p|_{\tilde x=-\tilde h/2}}{\tilde h},
\end{align}
the force per unit length in $x$-direction on top and bottom of the fiber $\tilde f_{\rm gap}$ can thus be calculated, using Eq.~(\ref{eq:px_av}), to
\begin{align}
\label{eq:2D_gap_force}
\tilde f_{\rm gap}(\tilde y) = -2\,\tilde h\,\bar \sigma_{xz}|_{\tilde z=b\tilde H} = -\frac{12\,\beta^2}{1-\beta^3}\,\bar u_p-\frac{\beta(1-\beta)}{2}\tilde H^2\Delta_{\partial_{\tilde x}\bar p},
\end{align}
with the pressure gradient variation
\begin{align}
\Delta_{\partial_{\tilde x}\bar p} = \partial_{\tilde x}\bar p\, - \langle\partial_{\tilde x}\bar p\rangle.
\end{align}
The factor $2$ in Eq.~(\ref{eq:2D_gap_force}) is needed to account for both the bottom and the top gap and the negative sign originates from the direction of the surface normal vector. The first term on the right-hand side of Eq.~(\ref{eq:2D_gap_force}) is identical to the gap force resulting from the model of \citet{Nagel2018}. The second term directly derives from the consideration of the variation of the flow velocity around its mean value in the gap. The composite particle velocity $\tilde u_p$ is determined by the condition of zero total force on the composite particle, which is equivalent to setting the total force on the fiber $\tilde F_{\rm tot}$ to zero, i.e.,
\begin{align}
\label{eq:zero_force_2D}
\tilde F_{\rm tot}=\int_{-1/2}^{1/2}d\tilde{y}\,\left(\tilde f_{\rm front}+\tilde f_{\rm tail}\right)+\tilde F_{\rm e}+\frac{12\,\beta^2}{1-\beta^3}\,\bar u_p = 0.
\end{align}
Only the first two terms on the left-hand side of Eq.~(\ref{eq:zero_force_2D}) are calculated from the numerical simulation, while the third term, i.e., the gap force, can be determined analytically once $\bar u_p$ is set. As for the 3D model, it is possible to exploit the linearity of the Brinkman equations to determine the composite particle velocity by the superposition of two reference configurations \cite{Bet2017}. The fiber velocity $\bar u_f$ is then directly given by Eq.~(\ref{eq:fib_vel_2D}).\\
It is important to note that the proposed extension of the gap flow is only valid for perpendicular fibers moving exclusively in $x$-direction. In the more general case of transversal motion, the $y$-dependence of the gap flow cannot be separated from the composite particle velocity in a straight-forward manner. For this reason, the contribution of the pressure gradient variation along the fiber on the total force distribution is evaluated in Appendix~\ref{app:comp_models}, so that possible errors by neglecting this effect can be estimated for future studies.
\subsection{Boundary conditions}
\label{sec:bc}
As already introduced in section \ref{sec:scaling_arg}, once the viscous and pressure forces are determined we derive the fiber deformation from the Euler-Bernoulli equation (\ref{eq:Euler_Bernoulli_scal}).  Because of the fiber being symmetric along $\tilde y$ at $\tilde y=0$, it is sufficient to consider half of it. The force applied on the two edges of the fiber $\tilde F_{\rm e}$ is taken into account via the boundary conditions,
\begin{align}
\label{eq:Euler_Bernoulli_BC}
\tilde\delta(\pm1/2)=0, \,\,\, \partial_{\tilde y}\tilde \delta(0)=0, \,\,\, \frac{1}{\tilde \mu}\,\partial_{\tilde y\tilde y}\tilde\delta(\pm 1/2) =\frac{\tilde F_{\rm e}}{4}, \,\,\, \frac{1}{\tilde\mu}\, \partial_{\tilde y\tilde y\tilde y}\tilde\delta(\pm 1/2)=\frac{\tilde F_{\rm e}}{2}.
\end{align}
The first condition arbitrarily sets the deflection to be zero at the edges as reference point. The second condition is due to the symmetry of the fiber and the last two conditions represent, respectively, the bending moment and the shear force due to the viscous forces applied on the edges of the fiber. The deflection is obtained by multiple numerical integration of one half of the obtained force distribution, leading directly to the maximum deflection $\tilde\delta_{\rm eq}$.
\subsection{Numerical implementation}
\label{sec:num_impl}
\begin{figure}
\centering
\includegraphics[width=\columnwidth]{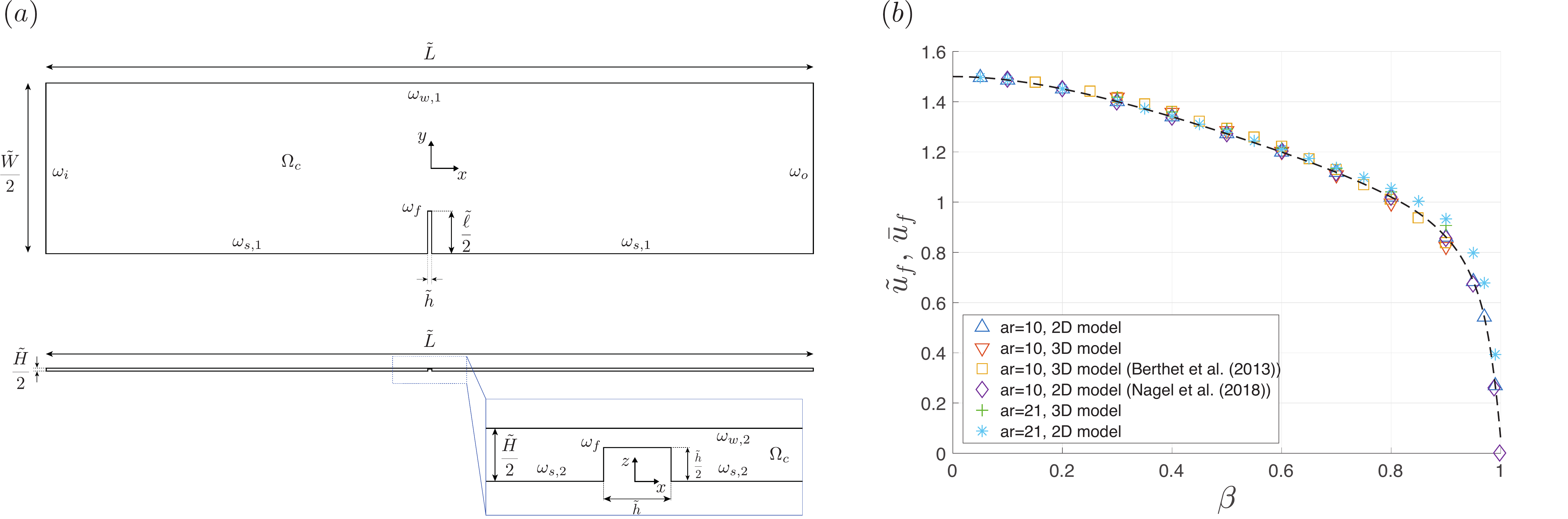}
\caption{Modeling: (a) Geometry of the 2D and 3D simulations: channel domain $\Omega_c$, lateral channel boundary $\omega_{w,1}$, symmetry plane $\omega_{s,1}$, inlet $\omega_i$, outlet $\omega_o$ and fiber surface $\omega_f$. Exploiting symmetry, only half of the domain in $y$- and $z$-direction is modeled. Top: Geometry of the 2D simulation and $xy$ plane view of the 3D simulation. Bottom: $xz$ plane view of the 3D simulation with additional top channel boundary $\omega_{w,2}$ and symmetry plane $\omega_{s,2}$.\\
(b) Validation against literature values: Dependency of the rigid fiber velocity $\tilde u_f$ on the confinement $\beta$ at force-free conditions for the 2D and 3D models for an aspect ratio $\rm ar=10$ and $\rm ar=21$ together with the data presented by \citet{Berthet2013} and \citet{Nagel2018}.}
\label{fig:geom_sim}
\end{figure}
Due to symmetry at $\tilde y=0$ and $\tilde z=0$, it is sufficient to simulate only a quarter of the channel in the 3D model and half of the channel in the 2D model. The fiber constitutes a hole in the channel. Figure~\ref{fig:geom_sim} shows the corresponding geometry and marks the channel domain $\Omega_c$ as well as the lateral and top channel boundaries, denoted by $\omega_{w,1}$ and $\omega_{w,2}$, respectively, the symmetry planes $\omega_{s,1}$ and $\omega_{s,2}$, the inlet and outlet, denoted by $\omega_i$ and $\omega_o$, respectively, and the fiber surface $\omega_f$. For both models, the dimensionless channel length and width are set to $\tilde L = 9$ and $\tilde W = 8$, while the remaining lengths are defined by Eqs.~(\ref{eq:scaled_lengths}).\\
We impose a slip condition on $\omega_{w,1}$, a no-slip condition on $\omega_{w,2}$ and a constant velocity, $\bar u_p$ in the 2D simulation and $\tilde u_f$ in the 3D simulation, on $\omega_f$. At the inlet $\omega_i$, a constant flow velocity $\tilde u_0\,\bs e_x$~(2D) or, respectively, a Poiseuille profile in $z$~(3D), i.e., $3/2\,\left(1-(2\,\tilde{z}\,\rm ar\,\beta)^2\right)\,u_0\,\bs e_x$, are prescribed, while the pressure is fixed to $\tilde p = 0$ at the outlet $\omega_o$ together with a normal outflow condition. The channel domain $\Omega_c$ is meshed with tetrahedral~(3D) or triangular~(2D) elements with maximum element size $\tilde H/2$~(3D) or $\tilde\ell/5$~(2D) and maximum element growth rate of $1.2$. In addition, the number of elements on the fiber edges is fixed to $m_h$ elements per $\tilde h$ and for the 3D simulation, the maximum element size on $\omega_f$ is fixed to $3\,m_h/\tilde h$, $m_h=60$ for the 2D model and $m_h=50$ for the 3D model, together with a maximum element growth rate of $1.2$. For both 2D and 3D simulations, the direct solver \textsc{pardiso} of \textsc{comsol multiphysics} software is utilized. Convergence for both models was checked with respect to $\tilde u_f$, or, respectively, $\bar u_f$, and $\tilde\delta_{\rm eq}$, details are given in App.~\ref{app:conv}.\\
\subsection{Model validation and assessment}
\label{subsec:modelvalidation}
We validate our models against the results from \citet{Berthet2013} and \citet{Nagel2018} by comparing the calculated fiber velocities for a rigid fiber as a function of the confinement $\beta$ for an aspect ratio $\rm ar = 10$, as shown by Fig.~\ref{fig:geom_sim}~(b).  The fiber velocity is observed to decrease monotonically with increasing confinement $\beta$ due to the increasing friction between the fiber and the top and bottom channel walls. Our results from both the 3D and the depth averaged 2D numerical simulations coincide very well with the existing results in literature, indicating the reliability of the present implementations. We have also displayed results for an aspect ratio of $21$, the value primarily used in the present work, and for $\beta>0.8$ they differ only slightly from the ones for $\rm ar=10$. In both configurations, the 2D model reproduces the results of the 3D model with high accuracy.\\
In the following we will use our models mainly to predict the force distribution $\tilde f$ and the fiber shape, including the maximum deflection $\tilde\delta_{\rm eq}$. The validity of our two models to predict these quantities will be discussed in detail in section \ref{sec:mech_fiber_def}, where we will show that both models lead to similar results.  As the 2D model offers a significantly higher flexibility and reduction of computational power and as the deviations from the 3D model are in the same range as the experimental error, we will primarily employ this model and provide additional results from the 3D~model when appropriate.
%
%
%
%
\section{Results}
\label{sec:results}

\subsection{Mechanism of fiber deformation}
\label{sec:mech_fiber_def}
Typical experimental observations can be seen in Fig. \ref{fig:reorientation} showing the successive positions and shapes of a flexible fiber during its transport along the microchannel.  Here we are interested in the deformed equilibrium shape of fibers of perpendicular orientation as indicated by the yellow arrows.  We investigate the maximum deflection $\delta_{\rm eq}$ as well as the corresponding fiber shapes as a function of different parameters as the mean velocity of the surrounding fluid $u_0$, the fiber length $\ell$, or the confinement $\beta$.

\begin{figure}
	\begin{center}
		\includegraphics[width=.9\columnwidth]{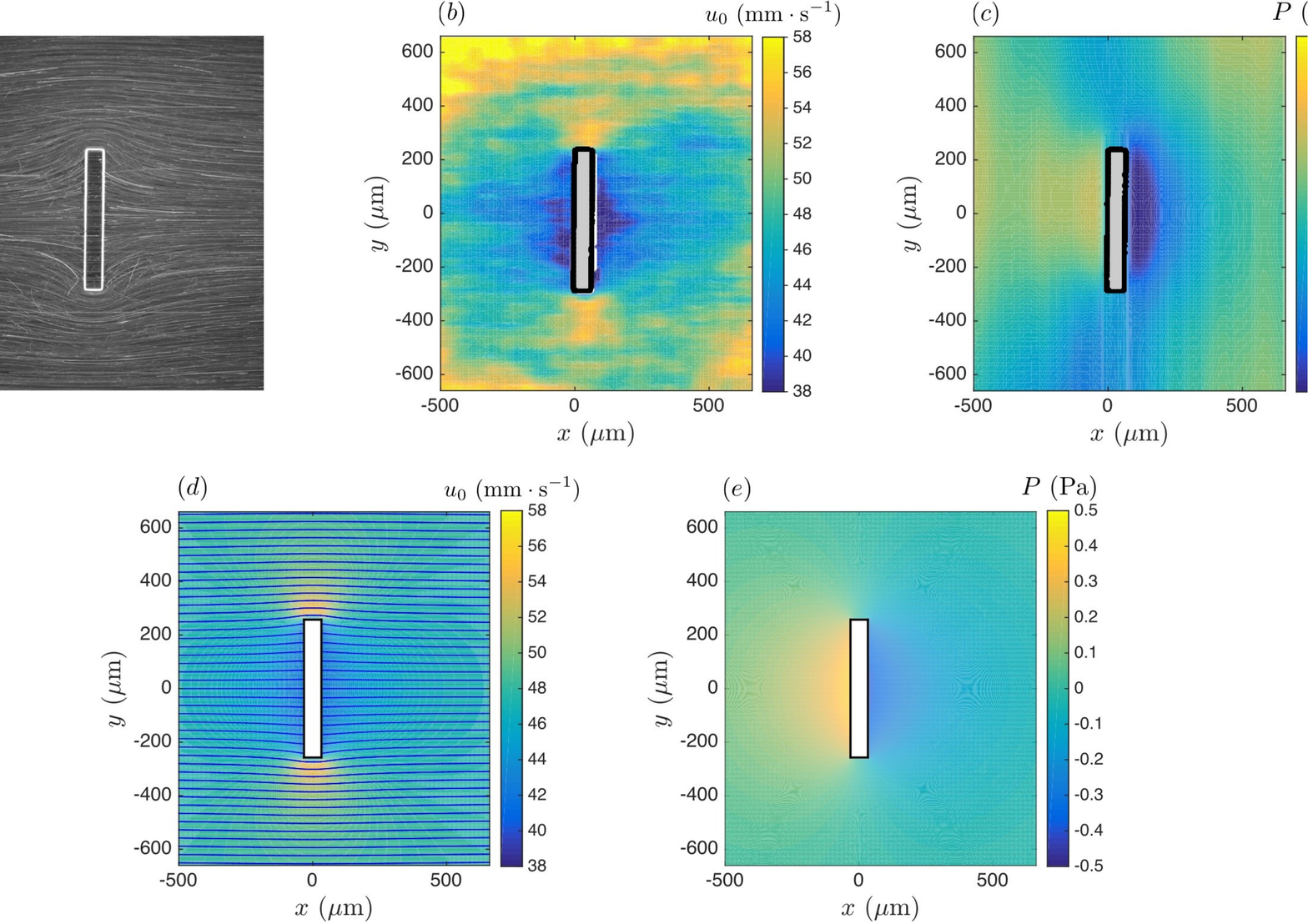}
	\end{center}
	\caption{Experiments: (a) Streaklines of the flow around a rigid fiber in the reference frame of the fiber. The mean fluid velocity is $u_0$ = 48 $\upmu$m$\cdot$s$^{-1}$, and the fiber dimensions are $\ell$ = 528 $\pm$ 5 $\upmu$m, $w$ = 67 $\pm$ 5$\upmu$m, and $h$ = 49 $\pm$ 3 $\upmu$m . Confinement is $\beta$ = 0.82. Streaklines are obtained by visualizing 1 $\upmu$m diameter beads flowing around the fiber with a X10 objective. Scale bar is 100 $\upmu$m. (b) Depth averaged velocity field around the fiber obtained from the particle tracking by averaging the particle velocity on time and on 64$\times$64 pixels windows. The constant velocity of the fiber has been added to get the velocity field in the frame of the laboratory. The noise on the edges of the window field results from the lack of data in this regions. (c) Pressure distribution minus the constant gradient in the ($Ox$) direction. It is obtained from the velocity field using the $2D$ Brinkman equation. Simulations: (d) Velocity field obtained numerically from the 2D model, the fiber dimension are length $L = 525 \upmu$m, width $w=65 \upmu$m and height $h=49 \upmu$m. Confinement is $\beta = 0.82$. In order to compare with experimental data the mean flow velocity is set to $48 \upmu$m$\cdot$s$^{-1}$. (e) Pressure field obtained from simulation with the same fiber geometry and confinement. The viscosity is set to $\mu = 67.4$ Pa$\cdot$s.  
	}
	\label{fig:streaklines}
\end{figure}

\begin{figure}[t]
	\centering
	\includegraphics[width=\columnwidth]{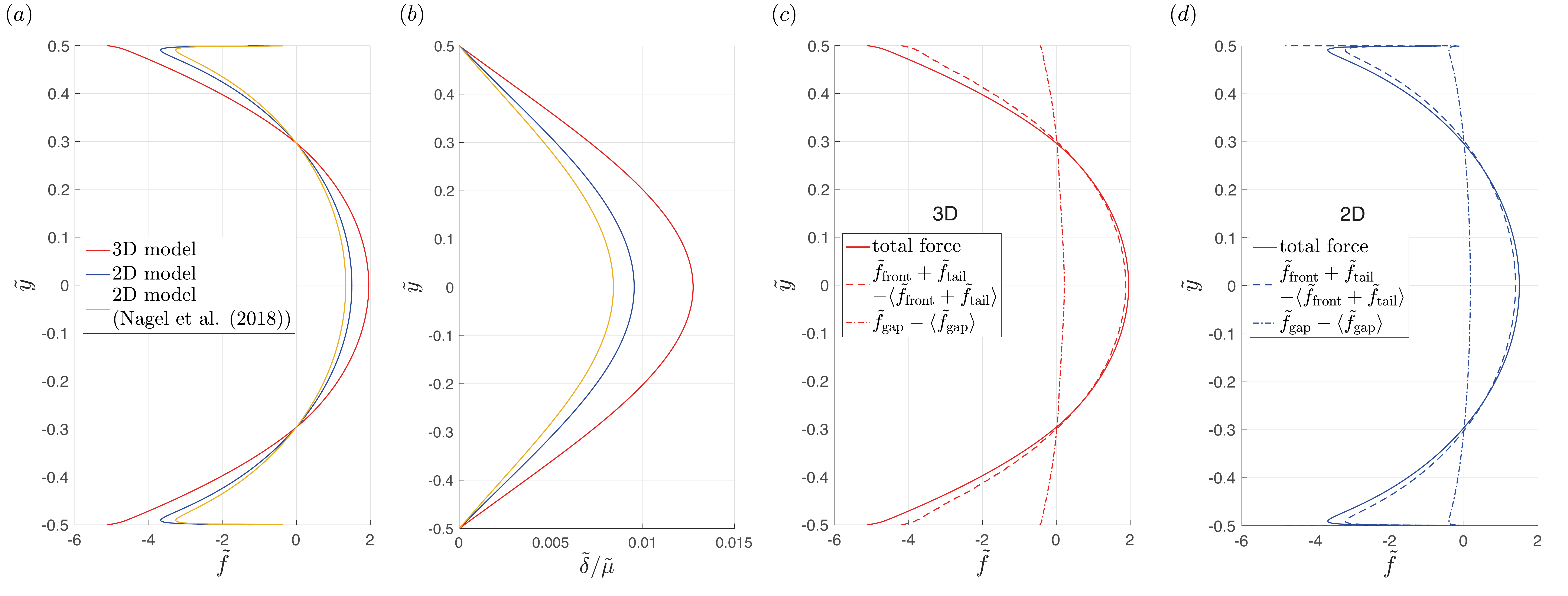}
	\caption{(a) Force distributions and (b) resulting normalized fiber deflections obtained from the 2D and 3D models. A smoothing function has been applied to the 3D data (see App.~\ref{app:edge_eff}). For comparison the results from the model by \citet{Nagel2018} have been added. The confinement is $\beta=0.8$ and the aspect ratio is $\rm ar = 21$. (c) and (d) Decomposition of the force distributions into the force on fiber front and tail and the force on fiber top and bottom for the 3D and 2D model, respectively. The mean values are subtracted for the sake of comparison.}
	\label{fig:comp_gap_force_models}
\end{figure}

The fiber deformation results from the inhomogeneous distribution of the drag force along the fiber due to its finite length. While being transported downstream, the flow pushes the fiber along the flow direction against the viscous friction, but it also flows around it. This leads to a specific flow profile and thus a specific pressure distribution around the fiber.  Fig. \ref{fig:streaklines} (a) shows an experimental visualization of the flow around a fiber in the reference frame of the fiber. Figure~\ref{fig:streaklines}~(b) shows the corresponding velocity field averaged over the channel height and Fig.~\ref{fig:streaklines}~(c) the pressure distribution. The pressure field is calculated from the experimental velocity field using the Brinkman equation. To distill the part of the pressure field causing a deformation of the fiber, we subtract the linear pressure field corresponding to a channel flow without fiber. The precise image and data treatment to obtain these results is described in Sec.~\ref{sec:exp_fib_transp}. Similar results are obtained from numerical simulations of the 2D model as shown by Fig.~\ref{fig:streaklines}~(d) and (e) for identical conditions. From these pictures two effects can be seen clearly: a non-homogeneous pressure distribution along the fiber length with a maximum pressure difference between the fiber front and tail located at the middle of the fiber, and an increase of the velocity close to the edges of the fiber.\\
From the non-homogeneous pressure distribution observed experimentally and numerically we expect a non-homogeneous force distribution along the fiber length. This force distribution is exemplarily shown in Fig.~\ref{fig:comp_gap_force_models}~(a) for $\beta=0.8$ and $\rm ar = 21$ for the 2D and 3D models. For the sake of completeness, the results from the model by \cite{Nagel2018} are shown as well. In all cases a maximum is observed at the center of the fiber, as indicated by the pressure distribution. It can also be seen that while all three models lead to similar results, the force is slightly underestimated by the 2D models. Our refined 2D model, however, leads to a correction towards the 3D results, and constitutes therefore an improvement compared to the model of \citet{Nagel2018}. \\

Figure~\ref{fig:comp_gap_force_models}~(b) shows the corresponding fiber shapes. Note that the non-zero shear force at the edges $\tilde F_e$ resulting from the increased flow velocity around the latter is taken into account through the boundary conditions~(\ref{eq:Euler_Bernoulli_BC}). All models lead to  the characteristic C-shape obtained in experiments, with a slightly varying amplitude for the different models. Figures~\ref{fig:comp_gap_force_models}~(c) and (d) show for the 3D and 2D model, respectively, the decomposition of the force distribution into the contributions of the pressure difference at fiber front and tail and the viscous shear forces on the top and bottom. The mean values (substracted from the force distributions in Figs.~\ref{fig:comp_gap_force_models}~(c) and (d)) of both $\tilde f_{\rm front}+\tilde f_{\rm tail}$ and $\tilde f_{\rm gap}$ are significant and almost equal in absolute value, as expected from the zero force condition and as the edge force $\tilde F_e$ is rather small. However, the variation around these mean values, which lead to the deflection of the fiber, is clearly more pronounced for the pressure difference than for the gap force. This reveals that the shape of the fiber is primarily determined by the pressure distributions and also comforts the use of an averaged 2D model. Note that the presence of sharp edges leads to fluctuations in the force distribution close to the edges. While this effect is clearly visible in the force distributions, the calculated deflections were not found to be significantly influenced by it. See appendix \ref{app:edge_eff} for details. In the following we will quantitatively discuss the fiber shape and maximum deflection from experiments and simulations as a function of the control parameters of the system.
\subsection{Fiber deflection as a function of the elasto-viscous number}
\label{subsec:elastovisc}
\begin{figure}
  \begin{center}
    \includegraphics[width=1\columnwidth]{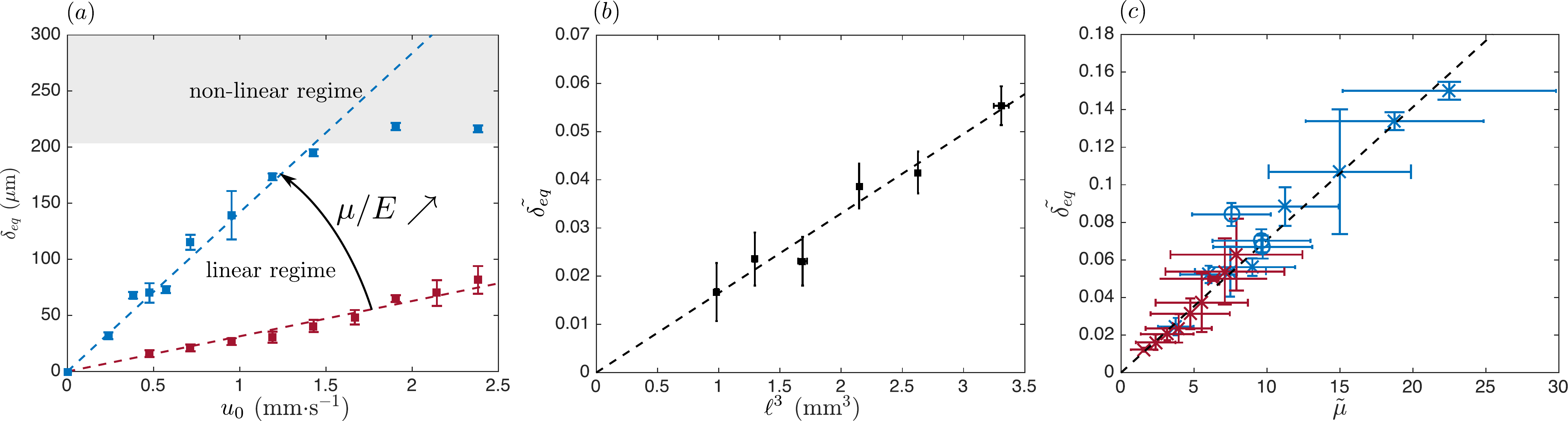}
  \end{center}
  \caption{Equilibrium deflection for a given confinement $\beta=0.8$. (a) Evolution of the equilibrium deflection as a function of the fluid mean velocity for a given fiber length $\ell =  1300 \upmu$m and for two different ratios $\mu/E = 3.14 \pm 0.08 \upmu$s (blue) and $\mu/E = 0.911 \pm 0.026 \upmu$s (red). (b) Evolution of the dimensionless equilibrium deflection as a function of fiber length to the power three for a given flow velocity $u_0=0.41$mm$\cdot \text{s}^{-1}$. Each point corresponds to two to four measurements. (c) Evolution of the dimensionless equilibrium deformation as a function of the elasto-viscous number $\tilde{\mu}$. The cross markers correspond to the data shown in (a) and circles are obtained with a smaller fiber, $\ell=  985 \pm 10 \upmu$m for fluid velocities varying from 0.46 mm$\cdot \text{s}^{-1}$ to 0.69 mm$\cdot \text{s}^{-1}$. The color code is the same as for (a) and distinguishes the different ratios $\mu/E$. Dotted lines correspond to linear fits.}
\label{fig:influence_vitesse}
\end{figure}
For small deformations we expect the scaled deformation $\tilde\delta_{\rm eq}$ to be proportional to the elasto-viscous number $\tilde{\mu}$, as given by Eq.~(\ref{eq:Euler_Bernoulli_scal}).  We first test the dependence on the fluid velocity by performing several sets of experiments keeping all other parameters constant. Figure~\ref{fig:influence_vitesse}~(a) illustrates that $\delta_{\rm eq}$ increases indeed linearly with the fluid velocity as long as the deformation remains small. For deformations larger than $20\%$, deviations from the linear behavior are observed. All the following experiments have been performed in the linear regime. Figure~\ref{fig:influence_vitesse} (a) also shows that with increasing ratio $\mu/E$, the deflection increases for identical flow velocities. Figure~\ref{fig:influence_vitesse}~(b) shows a series of experiments where the fiber length was modified and all other parameters were kept constant. As expected, $\tilde\delta_{eq}\propto \ell^3$. Note that the range of length available is small (the length varies from 990 $\pm$ 10 $\upmu$m to 1490 $\pm$ 10 $\upmu$m), as for too short fibers the deflection is too small to be measurable and for too long fibers the influence of the lateral channel walls can no longer be neglected (Sec.~\ref{sec:expmethods}).
Figure~\ref{fig:influence_vitesse}~(c) regroups experiments performed for different flow velocities, ratios $\mu/E$ and several fiber geometries for a constant confinement $\beta = 0.80 \pm 0.06$. It represents the scaled deflection as a function of $\tilde{\mu}$. The data points of Fig.~\ref{fig:influence_vitesse}~(c) align very well on a straight line, proving that $\tilde{\mu}$ indeed controls the amplitude of fiber deflection. Here, each point corresponds to the average value over several experiments and the large error bars are mainly due to the uncertainties on the determination of the Youngs modulus. While the Young modulus is accurately controlled, \textit{i.e.} similar fabrication conditions lead to identical fibers, its absolute value is determined with an error. The large error bar reflects this determination error but the alignment of the data points indicate the good accuracy in the mechanical properties. This remark is also true for all the figures representing data normalized by the elasto-viscous number.

%
\subsection{Effect of the confinement}
\label{sec:infl_conf_defl}
\begin{figure}
	\begin{center}
		\includegraphics[width=.8\columnwidth]{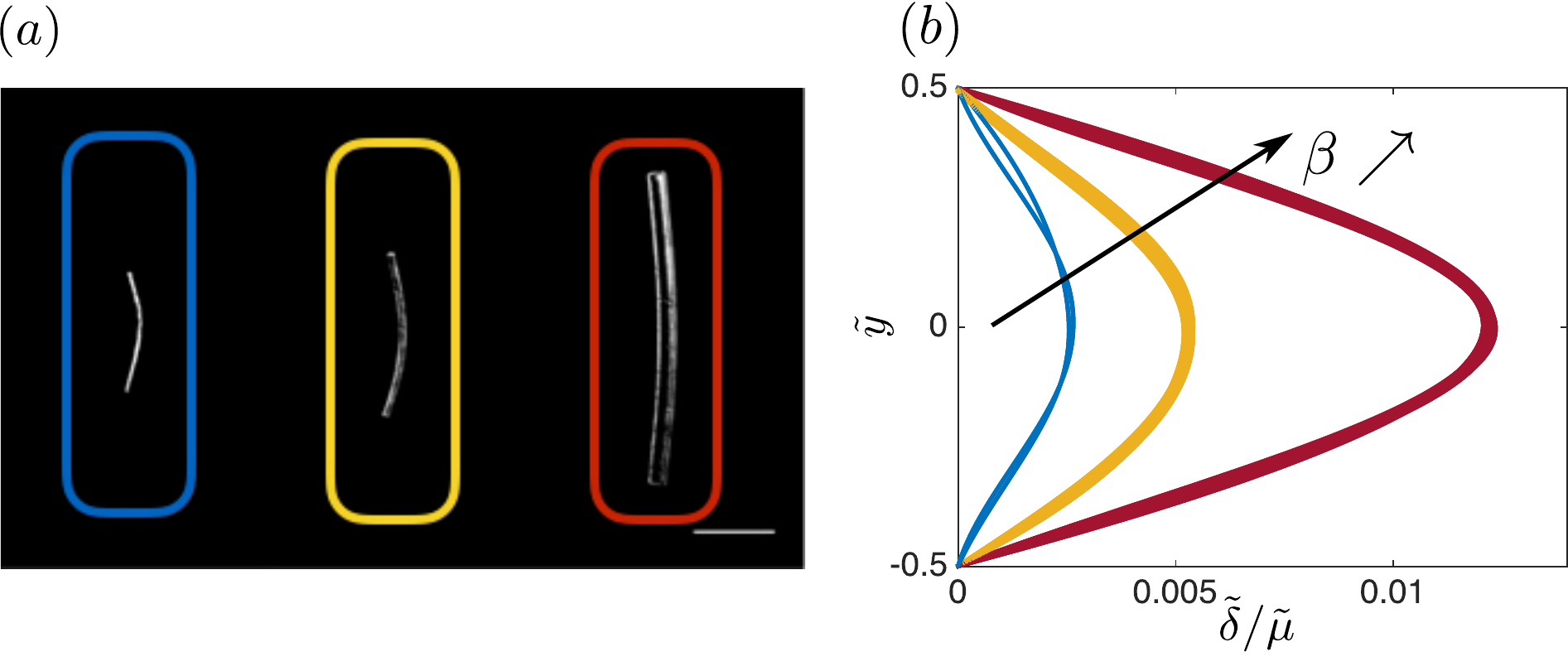}
	\end{center}
	\caption{Role of confinement. (a) Pictures of deformed fiber for confinements of $0.6 \pm 0.1$ (blue), $0.73 \pm 0.08$ (yellow) and $0.86 \pm 0.04$ (red). Scale bar is 500 $\upmu$m. (b) Dimensionless fiber shapes. $\tilde\delta$ has in addition been normalized by $\bar{\mu}$. The color code is the same as the for (a). Several fiber shapes for identical conditions have been superimposed.}
	\label{fig:influence_mu}
\end{figure}
Previous studies have shown that the confinement tunes the flow perturbation around the fiber and as a consequence the drag forces applied on the fiber \cite{Berthet2013, Nagel2018}. We thus expect the fiber deformation to strongly depend on the confinement. Fig.~\ref{fig:influence_mu}~(a) shows snapshots of fibers transported in channels of different confinement. A variation of the confinement implies a homothetic variation of the fiber width and length in order to keep a constant aspect ratio and a square cross-section. Despite the fact that the scaled deflection is normalized by $\tilde{\mu}$  in Fig.~\ref{fig:influence_mu}~(b), in order to compensate for the geometrical variations, different amplitudes of deflection are observed for different confinements. In the following we will discuss the influence of the confinement experimentally and numerically.\\

\begin{figure}
\begin{center}
\includegraphics[width=.7\columnwidth]{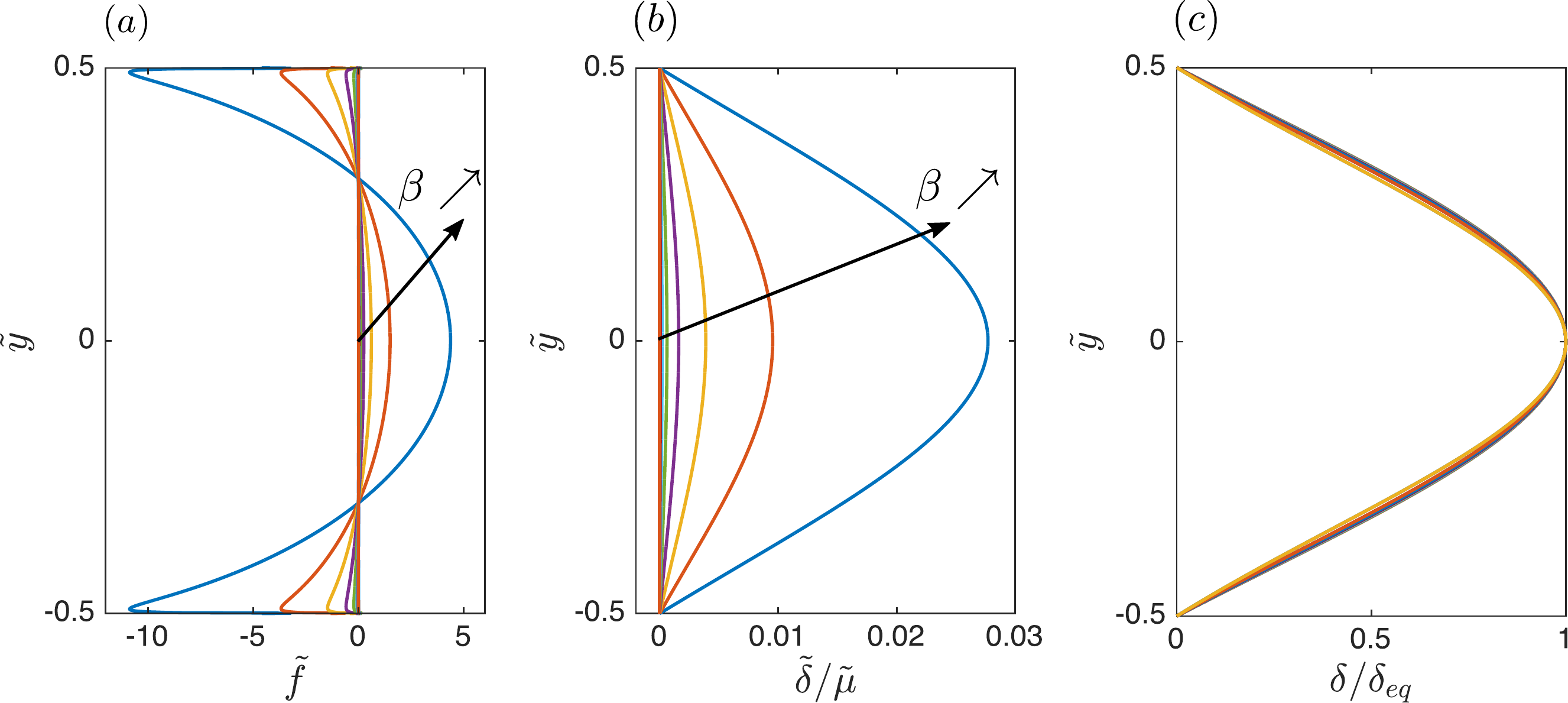}
\end{center}
\caption{(a) Evolution of the force distribution along the fiber length for confinements varying from $\beta=0.1$ to $\beta=0.9$ 
for a fiber with a squared cross section and an aspect ratio $\rm ar = 21$ transported in a confined channel without lateral walls. (b) Resulting normalized fiber deflections for the 2D model. (c) Fiber shapes normalized by $\delta_{eq}$ for different confinements.}
\label{fig:force_distrib}
\end{figure}

\begin{figure}
  \begin{center}
    \includegraphics[width=.7\columnwidth]{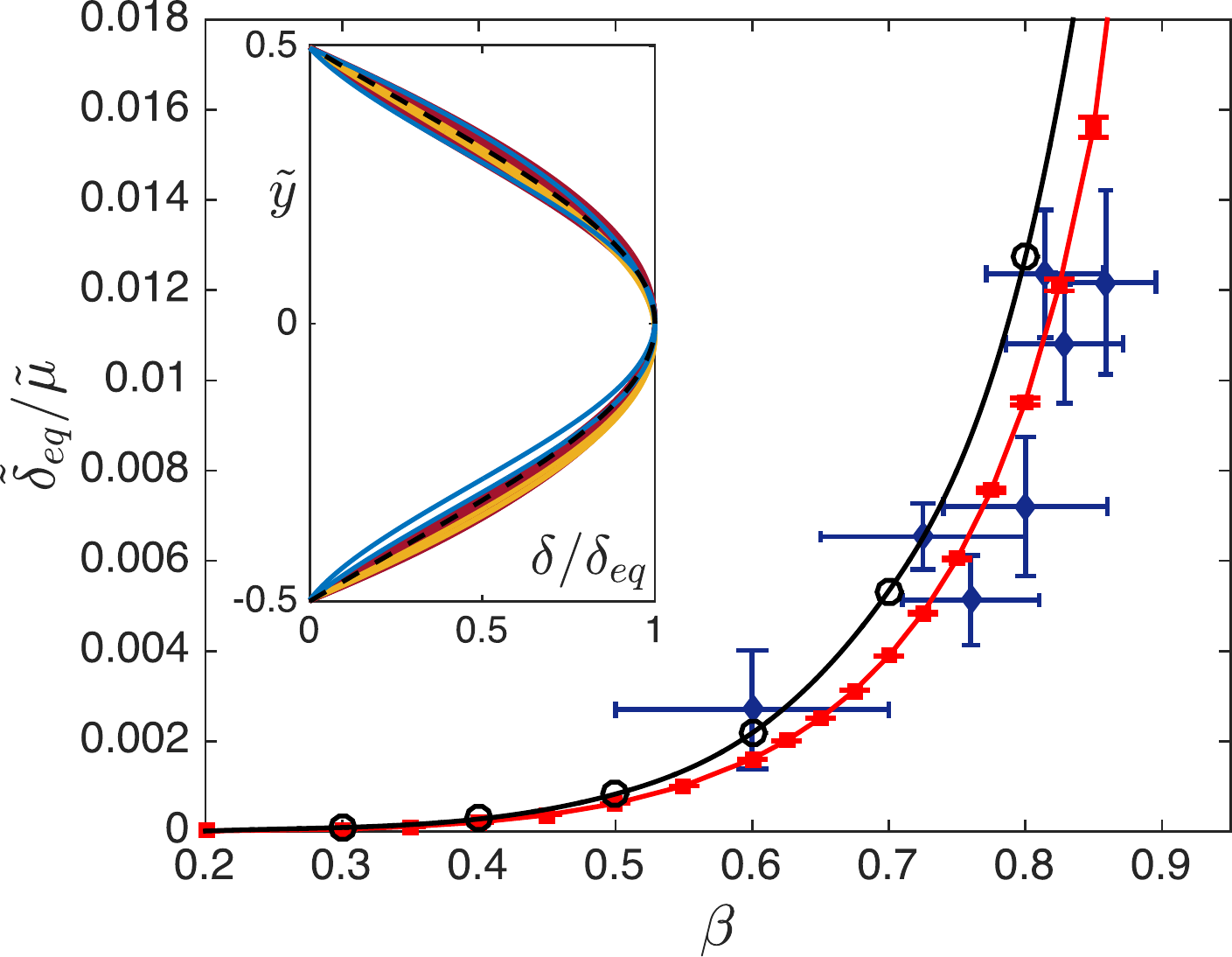} 
  \end{center}
  \caption{Evolution of the renormalized deflection of the fiber $\tilde{\delta}/\tilde{\mu}$ as a function of the confinement $\beta=h/H$ for a fiber of aspect ratio $\rm{ar}=21$. The blue diamonds show experimental data, each point corresponding to an average over the renormalized deflection for varying $\tilde{\mu}$. The red squares show the prediction of the $2D$ model and the black circles correspond to the 3D model. Red and black lines are guidelines. The inset shows experimental fiber shapes normalized by $\tilde\delta_{\rm eq}$ for different confinements (Blue: $\beta = 0.60 \pm 0.10$, yellow: $\beta = 0.73 \pm 0.08$, red: $\beta = 0.86 \pm 0.04$). The black dashed line indicates the shape obtained from the 2D numerical simulations for a confinement $\beta=0.8$.}
\label{fig:summary_beta}
\end{figure}

The force distribution along the fiber length obtained numerically using the 2D model is shown in Fig.~\ref{fig:force_distrib}~(a), for confinements varying from $\beta=0.1$ to $\beta=0.9$. The amplitude of $\tilde{f_{x}}$ increases with increasing confinement. Figure~\ref{fig:force_distrib}~(b) shows the related fiber shape and clearly indicates that the amplitude of fiber deflection also increases with confinement, as it was observed in experiments. The last panel of this figure shows the fiber shapes normalized by the maximum deflection, and reveals no significant change in the deflection shape. This means that the confinement mainly modifies the amplitude of the force and not its distribution, i.e., the fiber deflection is fully characterized by $\tilde\delta_{\rm eq}$. Similarly, normalized experimental fiber shapes for different confinements collapse onto a single shape, coinciding with the numerical result, as shown in the inset of Fig.~\ref{fig:summary_beta}.

Finally, we quantitatively discuss the amplitude of deflection $\tilde\delta_{\rm eq}$ as a function of the confinement by comparing experimental results to numerical calculations obtained from both 2D and 3D models. Figure~\ref{fig:summary_beta} superimposes the experimental (blue diamonds) and the numerical results obtained from the 2D (red squares) and the 3D (black circles) models. Each experimental point corresponds to 2--30 measurements for different fluid velocities, Youngs moduli and fiber dimensions.\\
The experimental results show good quantitative agreement with the numerical predictions. The results obtained from the 3D simulations predict slightly stronger deformations compared to the results obtained from the 2D model. A comprehensive comparison of the maximum deflection obtained from the two models and the one from \citet{Nagel2018} for various confinements is given in Appendix~\ref{app:comp_models}. In particular for high confinements, the results from the different models differ only slightly, confirming again the validity of the simplified 2D model which offers a significantly higher flexibility and reduction of computational power. In all cases, the differences in predictions of both models are in the same range as the experimental error and both are found to be in good agreement with the experimental data. \\
The deflection is observed to increase strongly with the confinement for $\beta \geq 0.6$. For less confined fibers, only a weak dependence can be observed. A similar observation had been made for the transport velocity of rigid fibers in confined channels (see Fig.~\ref{fig:geom_sim}), where for confinements above 0.6 a strong decrease of the transport velocity is observed, whereas for smaller confinements only a small influence is present.

%
%
\section{Conclusion and Outlook}
In this paper we have shown that flexible fibers transported in a plug flow deform when confined by the top and bottom walls. 
Fibers transported perpendicularly to the flow direction exhibit a C-shape, reflecting directly the nonuniform force distribution imposed on the fiber. The fiber acts as a moving obstacle when pushed by the flow against the friction with top and bottom walls along the channel. This leads to a perturbation of the flow and thus to non-homogeneous pressure and force distributions along the fiber, while the total force remains zero due to negligible inertia. While the variations in fiber deflection can be rationalized with an elasto-viscous number $\tilde{\mu}$, the force distribution strongly depends on the confinement as reflected by a sharp increase of the deflection with increasing confinement. Comparison of experimental results with numerical modeling using either a 2D model based on the Brinkman equations or a 3D model confirms this behavior quantitatively.\\
The interaction between deformation and transport of such flexible fibers leads to interesting dynamics and in particular to a reorientation towards an orientation parallel to the flow direction. The observed dynamics bear some similarities to the dynamics of sedimenting flexible fibers, which, however, reach a stable final position perpendicular to the direction of sedimentation. Understanding these dynamics is of importance for the controlled transport of flexible fibers in confined geometries, as for example in enhanced oil recovery or fiber optics. It could also provide insight in the deformation and transport of more complex deformable particles in confined geometries, as for example vesicles or red blood cells. In addition, as it was shown here that the deflection of the fiber can directly be linked to the force distribution, flexible fibers could be used as microfluidic sensors.
\begin{acknowledgments}
The European Research Council is acknowledged for funding the work through a consolidator grant (ERC PaDyFlow 682367). MB acknowledges the Deutsche Forschungsgemeinschaft (DFG) for financial support through a fellowship (BE 6681/1-1 403680998).
\end{acknowledgments}
\vspace{-4 mm}
%
%
%
\appendix
\section{Supplementary information on the 2D and 3D models}
\label{app:si_models}     
\subsection{Convergence of numerical simulations}
\label{app:conv}
\begin{figure}[h]
\centering
\includegraphics[width=\columnwidth]{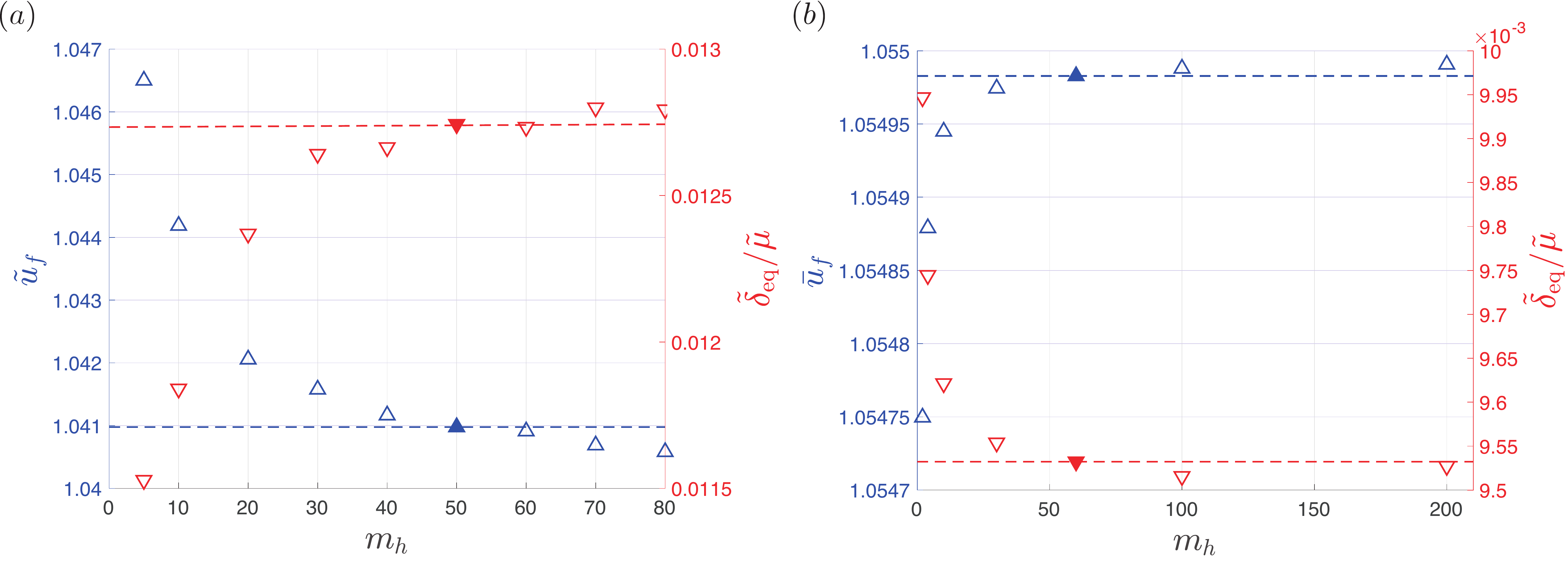}
\caption{Dependencies of fiber velocity $\tilde u_f$ or, respectively, $\bar u_f$, and renormalized maximum deflection $\tilde\delta_{\rm eq}/\tilde\mu$ on the mesh parameter $m_h$ for (a) the 3D model and (b) the 2D model with $\beta=0.8$ in both cases. The highlighted configuration is chosen for all calculations in this work and the dashed lines indicate the corresponding value for the sake of readability.}
\label{fig:mesh_conv}
\end{figure}

In order to ensure convergence of the numerical implementations presented in Sec.~\ref{sec:num_impl}, the influence of the mesh parameter $m_h$ on the fiber velocity and the normalized maximum deflection is analyzed for $\beta=0.8$ for both models and is depicted by Fig.~\ref{fig:mesh_conv}, while the resulting total number of mesh elements for each configuration is listed in Tab.~\ref{tab:mesh_elements}. A convergent behavior can be observed for both models with respect to the observed quantities. Especially in the case of the 3D model, the necessary computational power increases strongly with increasing number of mesh elements. Consequently, a compromise between efficiency and accuracy leads to $m_h=50$ for the 3D model and $m_h=60$ for the 2D model.
\begin{table}[h]
\centering
\setlength{\tabcolsep}{20pt}
\begin{tabular}{llll}
\hline
\multicolumn{2}{l}{3D model} & \multicolumn{2}{l}{2D model}\\
$m_h$ & total mesh elements & $m_h$ & total mesh elements\\
\hline
$5$ & $153844$ & $2$ & $3292$\\
$10$ & $180944$ & $4$ & $4185$\\
$20$ & $273681$& $10$ & $6932$\\
$30$ & $399679$& $24$ & $15916$\\
$40$ & $547532$& $\mathbf{60}$ & $\mathbf{29256}$\\
$\mathbf{50}$ & $\mathbf{734584}$& $100$ & $47150$\\
$60$	& $868090$& $250$ & $94448$\\
$70$ & $1168917$& & \\
$80$	& $1420441$& & \\
\hline
\end{tabular}
\caption{Total number of mesh elements in dependence of the mesh parameter~$m_h$ for both models. For the 3D model $\beta=0.8$, while the mesh is independent of the confinement for the 2D model. The chosen configuration is printed in bold and corresponds to the solid markers in Fig.~\ref{fig:mesh_conv}.}
\label{tab:mesh_elements}
\end{table}
\subsection{Consistency of the refined 2D model}
\label{app:consistency_2D_model}
\begin{figure}[h]
\centering
\includegraphics[width=0.5\columnwidth]{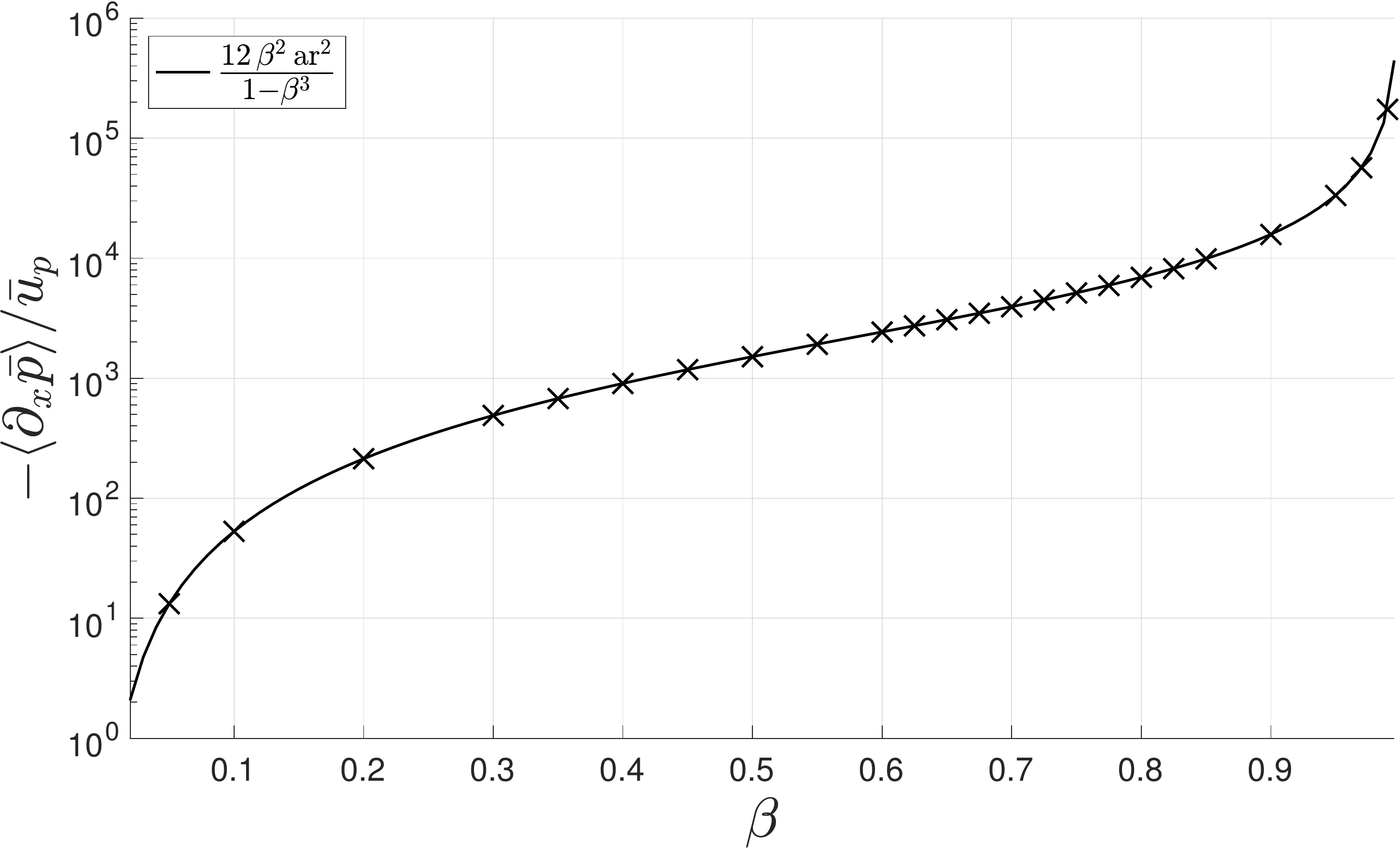}
\caption{Verification of condition (\ref{eq:px_av}). The single points are calculated based on the numerical simulation while the solid line depends only on $\beta$ and $\rm ar$, the latter being equal to $21$.}
\label{fig:gap_flow_ver}
\end{figure}

The extension of the gap flow to account for gap force variations along the fiber as it is presented in Sec.~\ref{sec:2D_model_eqs} implies a condition on the average of the pressure gradient $\langle\partial_xp\rangle$, which is given by Eq.~(\ref{eq:px_av}) and which should be checked in order to verify the consistency of the model with respect to the model presented by \citet{Nagel2018}, where only the mean value of the gap force is taken into account. For this reason, the pressure gradient is approximated according to Eq.~(\ref{eq:px_approx}), using the results of the numerical simulation, and numerically integrated. Figure~\ref{fig:gap_flow_ver} compares the values obtained by numerical calculation to the prescribed dependence on $\beta$ and reveals perfect coincidence.
\subsection{Comparison of maximum deflection predictions}
\label{app:comp_models}
\begin{figure}[h]
\centering
\includegraphics[width=0.8\columnwidth]{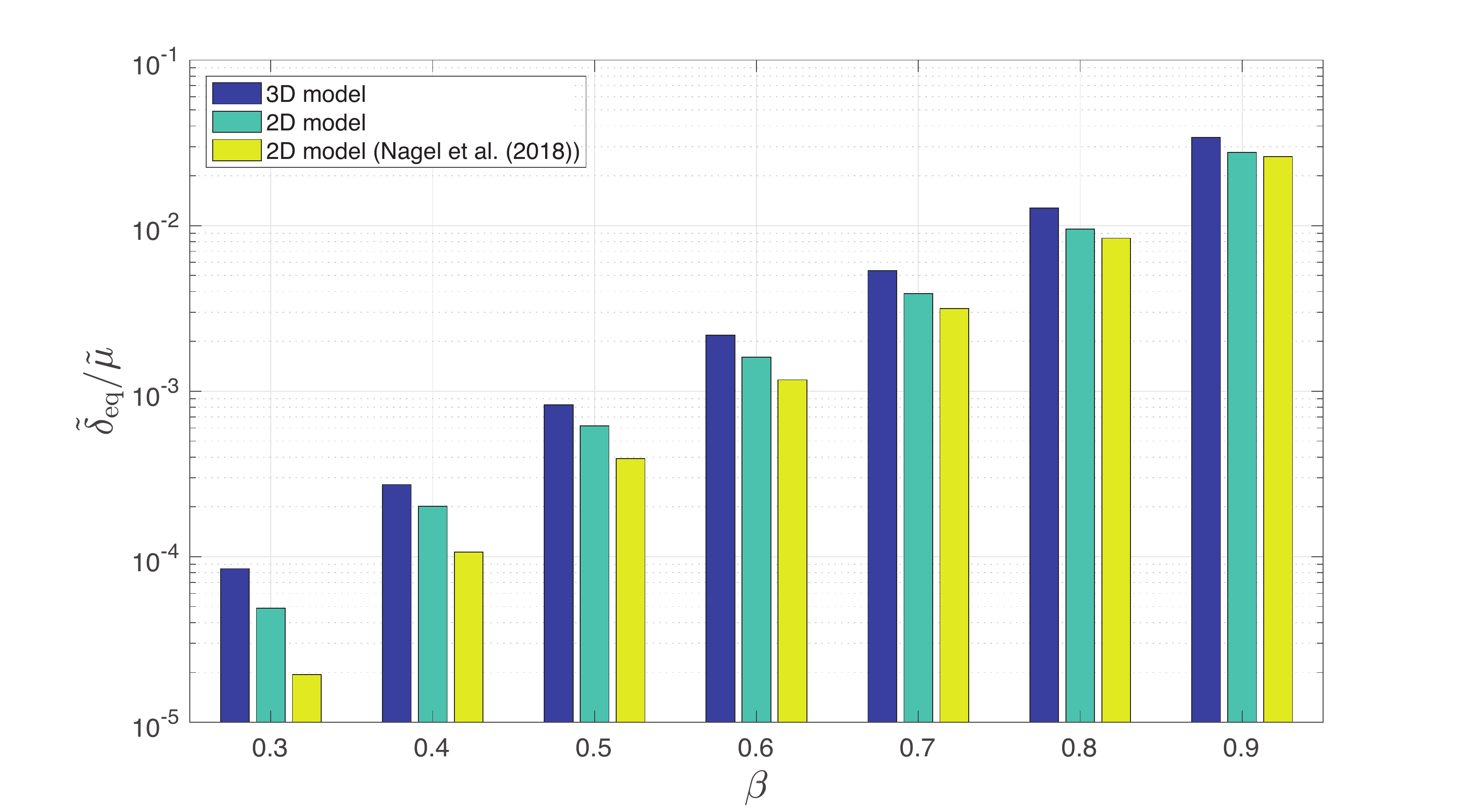}
\caption{Comparison of the normalized maximum deflection $\tilde\delta_{\rm eq}/\tilde\mu$ as calculated using the 3D and 2D models.}
\label{fig:comp_max_defl}
\end{figure}

Figure~\ref{fig:comp_max_defl} compares the determined maximum renormalized deflection for various values of the confinement~$\beta$ as obtained from the 3D and 2D models as well as from the 2D model from \citet{Nagel2018}, which can be obtained by setting $\Delta_{\partial_{\tilde x}\bar p}=0$ in Eq.~(\ref{eq:2D_gap_force}). While the 2D models systematically underestimate the maximum deflection, this effect decreases with increasing confinements. The difference between the two 2D models decreases as well with increasing confinements and is almost negligible for $\beta\geq0.8$.

For a perpendicular fiber the complexity of the calculations is not increased by including the gap force variations and they have thus been taken into account in the present work. However, the extension of the gap flow model is only possible for perpendicular fibers and approximating the gap force by its mean value as done by \citet{Nagel2018} can be advantageous for future studies on inclined fibers. 
\subsection{Edge effects}
\label{app:edge_eff}
\begin{figure}[h]
\centering
\includegraphics[width=0.6\columnwidth]{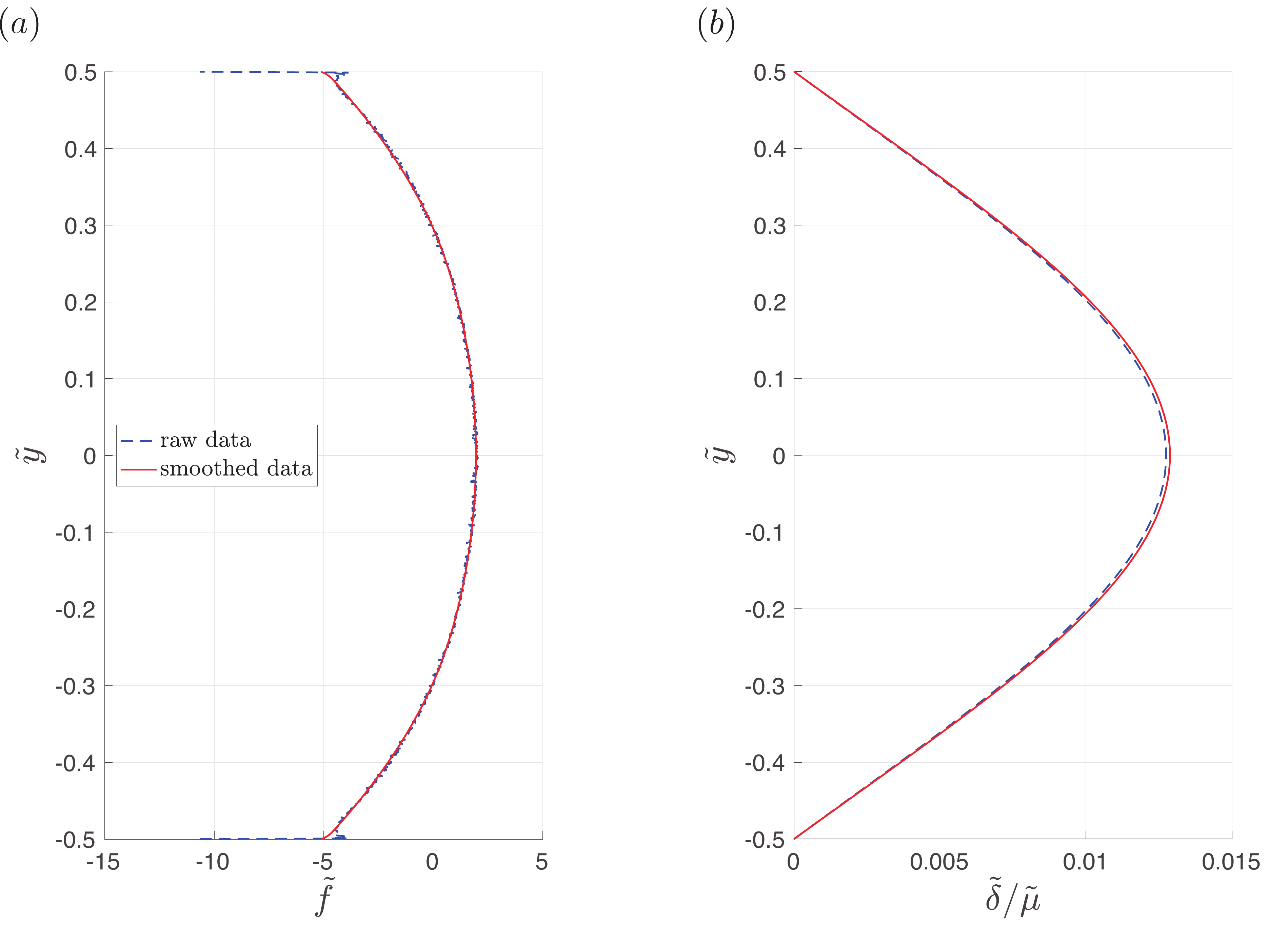}
\caption{Influence of smoothing the force distribution for the 3D model for $\beta=0.8$. (a) Raw and smoothed force distributions. (b) Resulting fiber deflections.}
\label{fig:raw_vs_smooth}
\end{figure}

Due to the sharp edges and corners of the fiber in the numerical implementations of both the 2D and the 3D model, fluctuations can be observed in the force distributions. For the 2D results, this can be seen at the edges of the force distribution in Fig.~\ref{fig:comp_gap_force_models}~(a) and (d), which seems to diverge. While this effect is clearly visible in the force distributions, the calculated deflections were not found to be significantly influenced by it. This is demonstrated in Fig.~\ref{fig:raw_vs_smooth} for the 3D model for $\beta=0.8$. Here, the force distribution appears to be scattered along the fiber, which is caused by averaging across the height and thus including the effect of the top and bottom edges. The magnitude of variation of the force around zero is rather small compared to the absolute forces at the fiber surfaces. As a consequence, small fluctuations can already lead to visible uncertainties in the data. We therefore smooth the data additionally, which implies truncating the apparent divergencies at $\tilde y =\pm 0.5$, as depicted by Fig.~\ref{fig:raw_vs_smooth}~(a). The resulting calculated deflections visualized by Fig.~\ref{fig:raw_vs_smooth}~(b) exhibit no significant differences.
%
%
%
%
%

%
\end{document}